\documentclass[a4paper]{article}
\pdfoutput=1

\usepackage{graphicx}
\usepackage{pstool}
\usepackage{amsmath}
\usepackage[numbers,sort&compress, elide]{natbib}
\usepackage{notes2bib}
\usepackage{verbatim}
\usepackage{sectsty}
\usepackage{setspace}
\usepackage[colorlinks=true, urlcolor=blue]{hyperref}
\usepackage{float}
\usepackage{chngcntr}
\usepackage{indentfirst}
\usepackage{authblk}
\usepackage{array}
\usepackage[capitalize]{cleveref}
\usepackage{ogonek}

\makeatletter
\def\@seccntformat#1{\csname the#1\endcsname.\quad}
\makeatother

\partfont{\fontsize{16pt}{20pt}\centering\selectfont}
\sectionfont{\fontsize{14pt}{18pt}\selectfont}
\subsectionfont{\fontsize{12pt}{16pt}\selectfont}

\newcommand{\A}{\text{\AA}}
\newcommand{\mtx}[1]{\boldsymbol{\hat{#1}}}
\newcommand{\bra}[1]{\left\langle #1 \right|}
\newcommand{\ket}[1]{\left| #1 \right\rangle}

\begin{document}

\title{A multi-scale approach to the electronic structure of doped semiconductor surfaces}
\date{}
\author[1]{Ofer Sinai}
\author[2]{Oliver T. Hofmann}
\author[2,3]{Patrick Rinke}
\author[2]{Matthias Scheffler}
\author[4]{Georg Heimel}
\author[1]{Leeor Kronik}

\affil[1]{Department of Materials and Interfaces, Weizmann Institute of Science, Rehovoth 76100, Israel}
\affil[2]{Fritz-Haber-Institut der Max-Planck-Gesellschaft, Faradayweg 4-6, D-14195 Berlin-Dahlem, Germany}
\affil[3]{COMP/Department of Applied Physics, Aalto University School of Science, P.O. Box 11100, FI-00076 Aalto, Finland}
\affil[4]{Institut f\"{u}r Physik, Humboldt-Universit\"{a}t zu Berlin, Newtonstr. 15, D-12489 Berlin, Germany}

\maketitle

\begin{abstract}
The electronic effects of semiconductor doping pose a unique challenge for first-principles simulations, because the typically low concentration of dopants would require intractably large system sizes in an explicit atomistic treatment. In systems which do not display long-range band bending, a satisfactory remedy is offered by the use of ``pseudoatoms'', with a fractional nuclear charge matching the bulk doping concentration. However, this alone is insufficient to account for charged semiconductor surfaces in the general case, where the associated space-charge region (SCR) may be very wide relative to tractable simulation dimensions. One generalization of the pseudoatom approach which overcomes this difficulty relies on the introduction of an artificially high doping level within a slab calculation, in conjunction with a multi-scale electrostatic energy correction. Here, we present an alternative technique that naturally extends the pseudoatom approach while bypassing the need for calculations with an unrealistically high doping level. It is based on the introduction of a two-dimensional sheet of charge within a slab-based surface calculation, which mimics the SCR-related field, along with free charge such that the system is neutral overall. The amount of charge involved is obtained from charge conservation and Fermi level equilibration between the bulk, treated semi-classically, and the electronic states of the slab/surface, which are treated quantum-mechanically. The method, which we call CREST - the Charge-Reservoir Electrostatic Sheet Technique - can be used with standard electronic structure codes. We validate the approach using a simple tight-binding model, which allows for comparison of its results with calculations encompassing the full SCR explicitly. Specifically, we show that CREST successfully predicts scenarios spanning the full range from no to full Fermi-level pinning. We then employ it with density functional theory, where it is used to obtain insights into the electronic structures of the ``clean-cleaved'' Si(111) surface and its buckled (2x1) reconstruction, at various doping densities.
\end{abstract}

\section{Introduction}
Semiconductor doping, which entails the introduction of impurities at concentrations typically of the order of parts per million, is of immense physical and technological importance. Beyond its extensive intentional use in semiconductor technology, many semiconductors (e.g., GaN \cite{VandeWalle1998} or ZnO \cite{Janotti2009}) often possess unintentional native defects that result in an effective doping which cannot be ignored. While such impurities/defects affect the chemical and crystalline properties of the bulk only locally, they dominate the electronic properties globally by increasing the amount of free charge carriers of either polarity, possibly by many orders of magnitude above the pristine semiconductor's intrinsic (thermally excited) concentration. This increase in charge carrier concentration is reflected in a shift of the Fermi level ($E_F$) position (typically) in the semiconductor forbidden gap \cite{Sze2006}.

At semiconductor surfaces (or interfaces), the existence of gap states may cause charge transfer between the doped material and its surface. The (typically) resulting depletion of free charge carriers from the region adjacent to the surface creates a space-charge region (SCR) originating from spatially fixed, ionized impurities. If the impurities are dilute, the SCR can extend hundreds of nm from the surface. The space charge creates an electric field which shifts the energy levels near the surface relative to their position in the neutral bulk, and thus relative to the global $E_F$, which is fixed by the bulk doping ratio and energy levels. For convenience, we use below the term ``surface Fermi level'' to indicate the position of $E_F$ relative to the band edges nearest the surface, which may differ from its position relative to the band edges in the neutral bulk (see, e.g., \cref{fgr:theory-main}(a)).

Over the years, understanding of various physically or technologically important phenomena has been obtained or enhanced by insights derived from first-principles computations in a wide range of systems involving semiconductors \cite{Cohen1988}, their surfaces \cite{Srivastava1997}, and their interfaces \cite{Capasso1987, Franciosi1996}. Density functional theory (DFT) is often the method of choice for such simulations \cite{Kaxiras2003, Martin2008} as it offers an unparalleled balance between accuracy and computational efficiency. In addition to the prediction of, e.g., system stabilities and reaction energies based on calculated total energies, critical insights into spectroscopic, magnetic and transport properties are obtained from the details of the electronic structure (see, e.g., Ref. \cite{Kronik2013}). However, the use of first-principles simulations imposes severe computational limitations on the ability to include global doping effects via the explicit insertion of dopants or defects. For example, a large unit cell may contain one thousand atoms. The replacement of one of these by a dopant atom corresponds to a doping concentration of the order of $10^{19}$ to $10^{20}$ cm$^{-3}$, depending on the material. However, practical doping concentrations are often lower than that by one to five orders of magnitude. Furthermore, in charged surface/interface scenarios, the width of the SCR often also precludes an explicit treatment.

Fortunately, microscopic phenomena at semiconductor surfaces/interfaces can often be decoupled from global doping effects. As a result, accurate microscopic descriptions of such systems may be obtained from first-principles simulations even though the bulk doping is ignored or its effects are accounted for separately. However, in many cases global doping effects may play a crucial part in determining the microscopic properties of the surface or interface. These include, to name a few experimental examples reported in recent literature, the observation of a doping-dependent surface dipole \cite{Yaffe2010, Yaffe2013}, control of the work function of ZnO surfaces via doping-dependent charge transfer to an adsorbed molecular acceptor \cite{Schlesinger2013, Winget2014}, enhancement of the photocurrent of a photosystem-1 film on Si \cite{LeBlanc2012}, and the doping-sensitivity of (2x4) reconstructions at the GaAs(001) surface \cite{Pashley1991} and (2x1) reconstructions at the Si(111) surface \cite{Bussetti2011, Feenstra2012}.

The first-principles simulation of doped semiconductor surfaces has been addressed in recent literature using different approaches. Krukowski, Kempisty, and co-workers have manipulated the location or charge of passivating, back-side atoms to reproduce the effects of the SCR on the electronic structure of finite slab simulations of charged surfaces \cite{Kempisty2009, *Kempisty2009e, Kempisty2011, *Kempisty2011e, Krukowski2013}. Using this, they demonstrated that the electric field of the SCR may shift surface state energies, a phenomenon that they termed the ``surface-state Stark effect''. However, it is not straightforward to relate the passivating-atom manipulations to the bulk doping, SCR field, and amount of charge transfer involved \cite{Krukowski2013}. Komsa and Pasquarello \cite{Komsa2013} have examined an \textit{a posteriori} correction scheme for the formation energy of charged defects at surfaces and interfaces. In this method, excess charge is explicitly introduced into slab-based surface calculations to mimic the free charge carriers introduced by doping. The total energies obtained from these calculations are then corrected for the spurious effects of the fixed, uniform counter-charge which accompanies the excess charge introduced. In both of the above approaches, a systematic way to determine the surface Fermi level in equilibrium with the bulk has not been offered.

Finally, work by some of us has demonstrated the successful incorporation of global doping effects in first-principles simulations of bulk, surfaces and interfaces \cite{Sinai2013, Moll2013, Richter2013, *Richter2013a, Xu2013, Hofmann2013}. In systems in which no SCR exists, or alternatively where the simulated semiconductor doping level is so high that the SCR width approaches tractable simulation dimensions, doping effects can be fully included in first-principles simulations, at no additional computational cost, solely by introducing fictitious ``pseudoatoms'' whose nuclear charge $Z$ (and corresponding electron number) is fractionally altered from the bulk atom integer number \cite{Sinai2013}. The added/subtracted fraction is set such that it corresponds to the bulk doping density of the semiconductor, leading to typically minute modifications of $Z$ from integer values. This approach can be thought of as a special case of the virtual crystal approximation (VCA). We refer here to such fractionally-charged atoms as ``doped'' atoms, so as to distinguish them from large-fraction pseudoatoms used in other contexts, e.g., for polar semiconductor surface passivation \cite{Huang2005} or to model alloys \cite{Jaros1985}. The use of such ``doped'' atoms eliminates the need to explicitly place a random distribution of dopants, leading to dramatically smaller computational cells. It was thus possible to reproduce the salient features of, e.g., a heavily-doped Si $p$-$n$ junction and fully passivated semiconductor surfaces \cite{Sinai2013, Hofmann2013}.

Despite this success, the use of stoichiometrically-correct ``doped'' atoms alone (VCA) is not sufficient to deal with systems displaying long-range band bending, e.g., charged surfaces of moderately-doped semiconductors. Because free charge is derived only from the ``doped'' atoms, the amount available for, e.g., transfer into a gap state is limited by the slab's actual size, causing the SCR formed to be spuriously confined in most cases. In other words, if the bulk represents a reservoir of charge at a certain chemical potential (the Fermi energy), then the ``doped-atom'' approach provides the means to adjust this energy, but it cannot always act as a sufficiently large reservoir. To deal with such systems, some of us have previously applied a generalized VCA-based method, using pseudoatom $Z$ modifications large enough to provide all of the surplus charge needed at the simulated surface within the slab supercell \cite{Richter2013, *Richter2013a, Moll2013, Xu2013}. This was equivalent to using the VCA for a material whose bulk doping density was artificially very high, such that the entire SCR would be contained inside the (nm-scale) thin slab. Purely electrostatic descriptions of the spuriously thin SCR and the real system's SCR were then used to correct the supercell's total energy \textit{a posteriori}. This ``multi-scale VCA'' approach was first applied to calculate the concentrations of neutral and charged oxygen vacancies at the surface of MgO(100) \cite{Richter2013, *Richter2013a}. It was then adjusted to examine the adsorption of hydrogen \cite{Moll2013} and of an organic molecule \cite{Xu2013} on ZnO($000\overline{1}$). In this approach the microscopic system calculated in practice was considerably perturbed from its actual (moderately-doped) state, due to both the electronic over-charging (at some stages of the treatment) and the over-modification of the pseudoatom nuclei. Thus, in addition to the electrostatic \textit{a posteriori} correction, derivation of correct total energies required extrapolation to low defect concentrations from several progressively larger surface supercell calculations \cite{Richter2013} or the introduction of an electronic ``filling'' correction \cite{Xu2013}. The subsequent derivation of the equilibrium surface charge for a given bulk doping concentration involved a numerical Gibbs' free-energy minimization scheme based on the extrapolated data \cite{Richter2013}, or the extrapolation of results from several calculations with arbitrary magnitudes of charge-transfer \cite{Xu2013}.

In cases where one wishes to study the effect of doping on the electronic structure itself, or more generally when one wishes to avoid extrapolation schemes from an artificially high doping level, it would be advantageous to use a technique which retains the system's stoichiometric doping level within the atomistic simulation. In this article, we present such a general scheme. We handle the problem of an arbitrarily-doped semiconductor surface at two inter-dependent levels of theory: The behavior of the surface is treated quantum-mechanically, while the bulk material is treated via the effective mass model. The two scales are linked by the twin requirements of Fermi-level equilibration and charge conservation in the entire system. This is achieved by using ``doped'' atoms representing the realistic bulk doping concentration within a slab-based surface calculation, complemented by the introduction of a two-dimensional sheet of charge which mimics the electrostatic field associated with the SCR. The charged sheet, similar to that employed in the past in the context of charged periodic calculations \cite{Lozovoi2003, Gava2009}, is placed in the vacuum outside the virtual, passivated face of the non-symmetric slab. We ascertain that it introduces no spurious effects at the slab's virtual face and in the system in general. Because the sheet is introduced along with free counter-charge, mimicking charge-transfer from the SCR, the overall neutrality of the system is preserved, avoiding spurious effects associated with charged periodic slab calculations. We refer to this technique as the Charge-Reservoir Electrostatic Sheet Technique, or CREST.

In \cref{sec:theory} of this paper, we explain CREST in detail. In \cref{sec:TB}, using CREST as a wrapper program around a simple tight-binding electronic-structure code, we validate its results by comparison to calculations in which the full SCR is included, and test it in a wide range of surface-defect and bulk-doping scenarios. Finally, in \cref{sec:DFT} we apply the method to DFT calculations of Si(111) surfaces, comparing at different doping densities the ``clean-cleaved'', unpassivated surface with the buckled Pandey-chain (2x1) reconstructions of this surface.

\section{Theory}
\label{sec:theory}
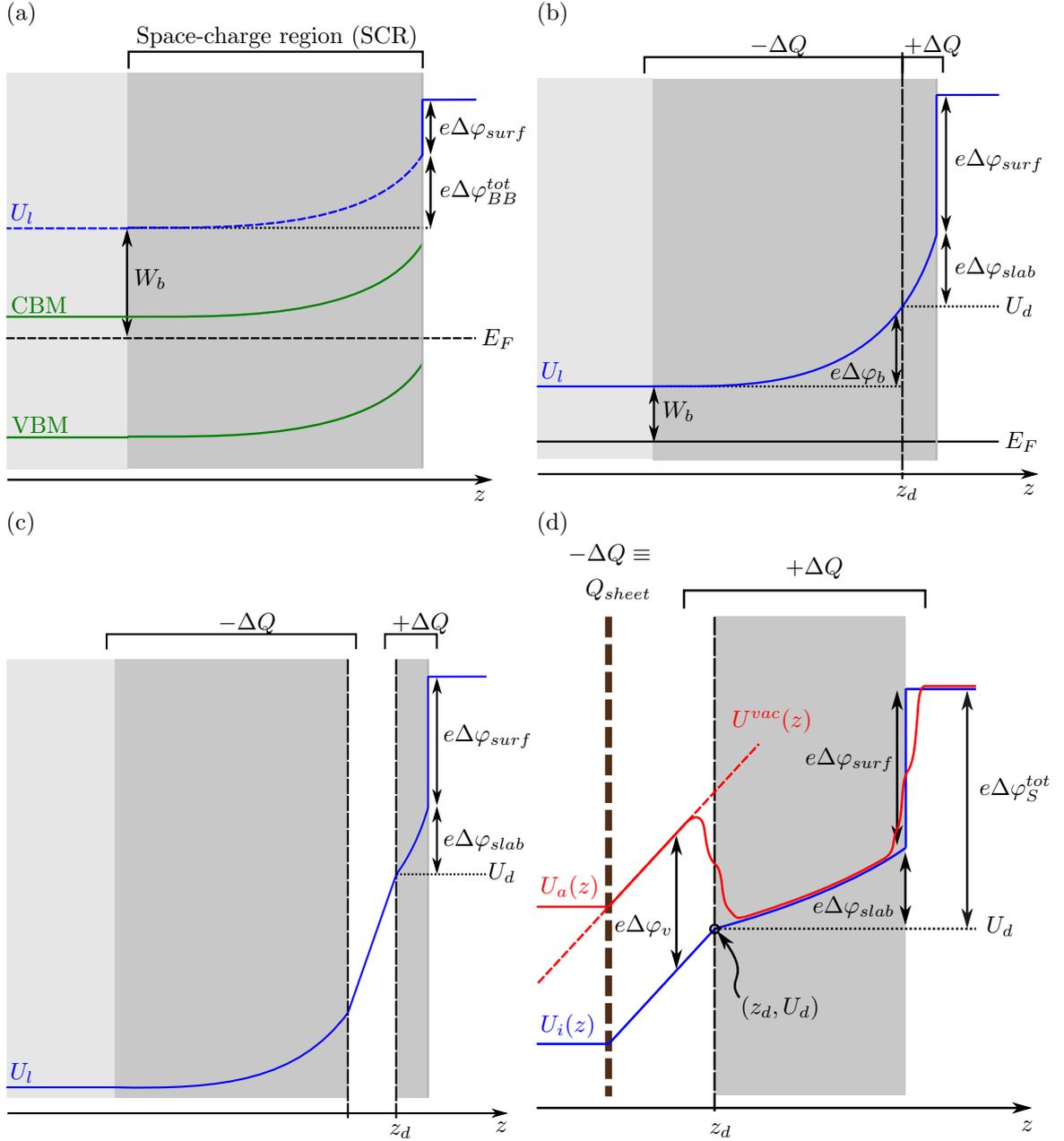
\begin{figure}[htb!]
\centering
  \begin{tabular}{ l l }
  (a) & (b)\\
\begingroup%
  \setlength{\unitlength}{0.5\textwidth}%
  \begin{picture}(1,0.94790402)%
    \put(0,0){\includegraphics[width=\unitlength]{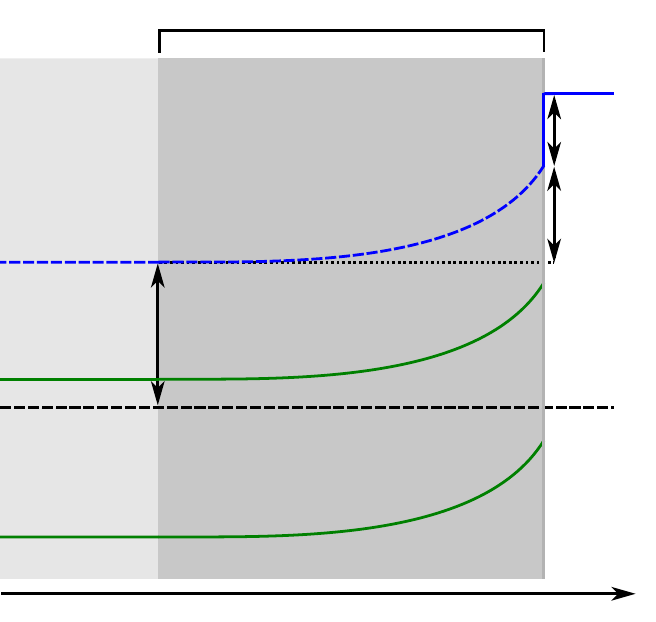}}%
    \put(0.86409385,0.60964739){\color[rgb]{0,0,0}\makebox(0,0)[lb]{\smash{$e\Delta\varphi_{BB}^{tot}$}}}%
    \put(0.86247364,0.73514626){\color[rgb]{0,0,0}\makebox(0,0)[lb]{\smash{$e\Delta\varphi_{surf}$}}}%
    \put(0.2545515,0.4318482){\color[rgb]{0,0,0}\makebox(0,0)[lb]{\smash{$W_b$}}}%
    \put(0.94727549,0.3127185){\color[rgb]{0,0,0}\makebox(0,0)[lb]{\smash{$E_F$}}}%
    \put(0.00832124,0.56319796){\color[rgb]{0,0,1}\makebox(0,0)[lb]{\smash{$U_l$}}}%
    \put(0.00945303,0.37695465){\color[rgb]{0,0.50196078,0}\makebox(0,0)[lb]{\smash{CBM}}}%
    \put(0.01123699,0.13685176){\color[rgb]{0,0.50196078,0}\makebox(0,0)[lb]{\smash{VBM}}}%
    \put(0.54002132,0.91930997){\color[rgb]{0,0,0}\makebox(0,0)[cc]{\smash{Space-charge region (SCR)}}}%
    \put(0.93150465,-0){\color[rgb]{0,0,0}\makebox(0,0)[lb]{\smash{$z$}}}%
  \end{picture}%
\endgroup%
  &
\begingroup%
  \setlength{\unitlength}{0.5\textwidth}%
  \begin{picture}(1,0.92779301)%
    \put(0,0){\includegraphics[width=\unitlength]{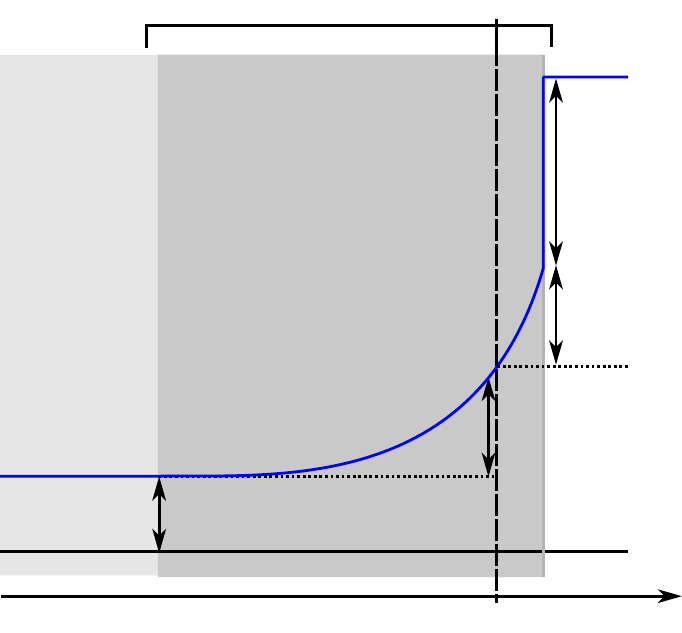}}%
    \put(0.69677263,0.24438448){\color[rgb]{0,0,0}\makebox(0,0)[rb]{\smash{$e\Delta\varphi_b$}}}%
    \put(0.83298198,0.67333506){\color[rgb]{0,0,0}\makebox(0,0)[lb]{\smash{$e\Delta\varphi_{surf}$}}}%
    \put(0.24986209,0.1668903){\color[rgb]{0,0,0}\makebox(0,0)[lb]{\smash{$W_b$}}}%
    \put(0.93418193,0.37719028){\color[rgb]{0,0,0}\makebox(0,0)[lb]{\smash{$U_d$}}}%
    \put(0.93311514,0.10474283){\color[rgb]{0,0,0}\makebox(0,0)[lb]{\smash{$E_F$}}}%
    \put(0.83298198,0.45614805){\color[rgb]{0,0,0}\makebox(0,0)[lb]{\smash{$e\Delta\varphi_{slab}$}}}%
    \put(0.00807033,0.24625544){\color[rgb]{0,0,1}\makebox(0,0)[lb]{\smash{$U_l$}}}%
    \put(0.42470242,0.90028878){\color[rgb]{0,0,0}\makebox(0,0)[lb]{\smash{$-\Delta Q$}}}%
    \put(0.73200819,0.90028878){\color[rgb]{0,0,0}\makebox(0,0)[lb]{\smash{$+\Delta Q$}}}%
    \put(0.96857753,0.01308384){\color[rgb]{0,0,0}\makebox(0,0)[lb]{\smash{$z$}}}%
    \put(0.71497289,0.00783834){\color[rgb]{0,0,0}\makebox(0,0)[lb]{\smash{$z_d$}}}%
  \end{picture}%
\endgroup%
  \\
  (c) & (d)\\
\begingroup%
  \setlength{\unitlength}{0.5\textwidth}%
  \begin{picture}(1,1.04790663)%
    \put(0,0){\includegraphics[width=\unitlength]{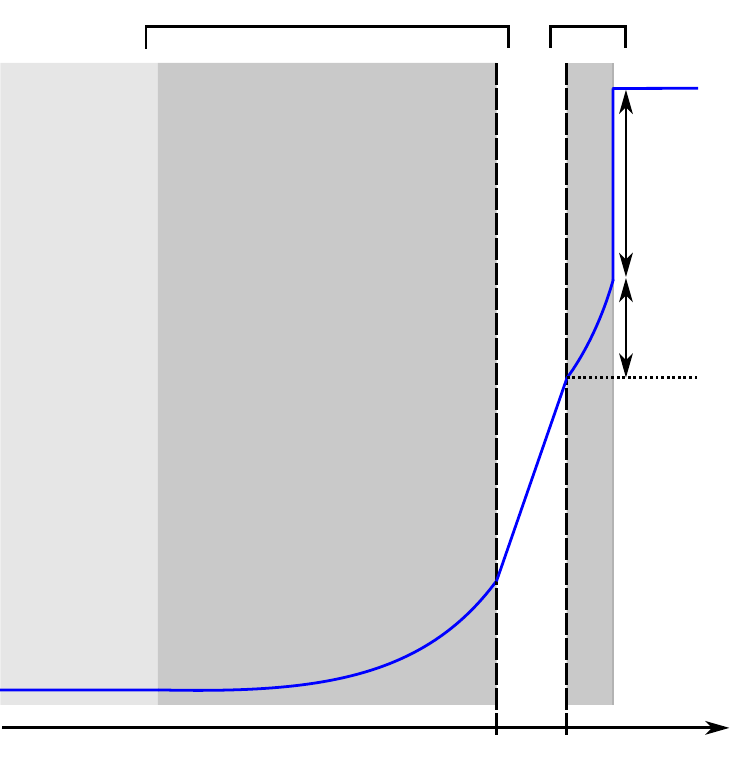}}%
    \put(0.87116449,0.79673748){\color[rgb]{0,0,0}\makebox(0,0)[lb]{\smash{$e\Delta\varphi_{surf}$}}}%
    \put(0.95995829,0.52006384){\color[rgb]{0,0,0}\makebox(0,0)[lb]{\smash{$U_d$}}}%
    \put(0.87116449,0.58753143){\color[rgb]{0,0,0}\makebox(0,0)[lb]{\smash{$e\Delta\varphi_{slab}$}}}%
    \put(0.00754844,0.11770146){\color[rgb]{0,0,1}\makebox(0,0)[lb]{\smash{$U_l$}}}%
    \put(0.42301532,1.02219313){\color[rgb]{0,0,0}\makebox(0,0)[lb]{\smash{$-\Delta Q$}}}%
    \put(0.76856798,1.02219313){\color[rgb]{0,0,0}\makebox(0,0)[lb]{\smash{$+\Delta Q$}}}%
    \put(0.97488969,0.01223292){\color[rgb]{0,0,0}\makebox(0,0)[lb]{\smash{$z$}}}%
    \put(0.7634953,0.00732801){\color[rgb]{0,0,0}\makebox(0,0)[lb]{\smash{$z_d$}}}%
  \end{picture}%
\endgroup%
  &
\begingroup%
  \setlength{\unitlength}{0.5\textwidth}%
  \begin{picture}(1,1.20570456)%
    \put(0,0){\includegraphics[width=\unitlength]{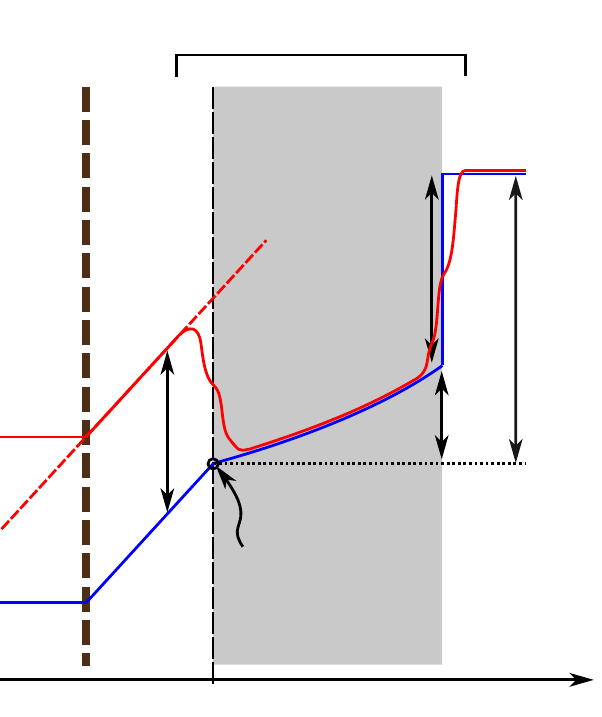}}%
    \put(0.53500000,0.75000000){\color[rgb]{0,0,0}\makebox(0,0)[lb]{\smash{$e\Delta\varphi_{surf}$}}}%
    \put(0.89573928,0.41342234){\color[rgb]{0,0,0}\makebox(0,0)[lb]{\smash{$U_d$}}}%
    \put(0.55003040,0.45241753){\color[rgb]{0,0,0}\makebox(0,0)[lb]{\smash{$e\Delta\varphi_{slab}$}}}%
    \put(0.49616857,1.13241606){\color[rgb]{0,0,0}\makebox(0,0)[lb]{\smash{$+\Delta Q$}}}%
    \put(0.14136154,1.12413694){\color[rgb]{0,0,0}\makebox(0,0)[b]{\smash{
    $\begin{aligned}
    -\Delta Q \equiv\\
    Q_{sheet}
    \end{aligned}$
    }}}%
    \put(0.96541322,0.01501795){\color[rgb]{0,0,0}\makebox(0,0)[lb]{\smash{$z$}}}%
    \put(0.34378019,0.00899635){\color[rgb]{0,0,0}\makebox(0,0)[lb]{\smash{$z_d$}}}%
    \put(0.00740612,0.21292360){\color[rgb]{0,0,1}\makebox(0,0)[lb]{\smash{$U_i(z)$}}}%
    \put(0.15498667,0.41399870){\color[rgb]{0,0,0}\makebox(0,0)[lb]{\smash{$e\Delta\varphi_v$}}}%
    \put(0.88175980,0.68899696){\color[rgb]{0,0,0}\makebox(0,0)[lb]{\smash{$e\Delta\varphi_S^{tot}$}}}%
    \put(0.00740612,0.49228055){\color[rgb]{1,0,0}\makebox(0,0)[lb]{\smash{$U_a(z)$}}}%
    \put(0.40597107,0.24868680){\color[rgb]{0,0,0}\makebox(0,0)[lb]{\smash{$(z_d,U_d)$}}}%
    \put(0.38884333,0.82560243){\color[rgb]{1,0,0}\makebox(0,0)[lb]{\smash{$U^{vac}(z)$}}}%
  \end{picture}%
\endgroup%
  \\
  \end{tabular}
  \caption{Explanation of the Charge-Reservoir Electrostatic Sheet Technique (CREST). See text for details and explanation of terms. (a) Schematic description of a SCR at a doped semiconductor surface. (b) The physical system is bisected at $z_d$ into two oppositely charged parts. (c) The semi-infinite part of the system is ``moved left'', creating a parallel capacitor with an intermediate vacuum ``layer''. (d) Description of a typical slab system after contraction of the SCR to a charged sheet. The idealized $U_l$ or electrostatic energy is shown as a blue curve, denoted by $U_i(z)$, and $U_l$ resulting from macroscopic averaging (see page \pageref{pp:Uaz}) of the electrostatic energy from a typical DFT or other atomistic calculation is shown as a red curve, denoted by $U_a(z)$ (slightly offset from $U_i(z)$ to make it easier to differentiate). Also shown is the linear curve extrapolated from $U_a(z)$ in the vacuum between the charged sheet and the slab's virtual surface ($U^{vac}(z)$, red dashed line).}
  \label{fgr:theory-main}
\end{figure}

The general physical system which we wish to simulate is a semi-infinite slab with a surface, at which surface gap-states exist. This causes charge transfer to/from the surface, creating a SCR in the semiconductor, wherein band-bending occurs. Such a system is shown schematically in \Cref{fgr:theory-main}(a). In this figure, the surface normal lies along the z-axis, with the semi-infinite bulk extending to the $z \rightarrow -\infty$ direction (left). The system is infinite in the plane of the surface. The neutral bulk is shaded light gray and the SCR is shaded dark gray. The conduction band minimum (CBM), valence band maximum (VBM), and Fermi level ($E_F$) are marked on the figure. Importantly, we also mark on the figure the local vacuum level, $U_l$, defined as the energy of an electron at a certain point at rest, if the periodic crystal potential is removed \cite{Marshak1989}. Practically, $U_l$ represents the potential energy due to effects other than the crystal potential, including, e.g., the SCR and the surface dipole \cite{Kronik1999}. Accordingly, within the bulk $U_l$ is parallel to the semiconductor band edges (and higher relative to CBM by an effective electron affinity), whereas in the vacuum far outside the surface (where no crystal potential is present), $U_l$ coincides with the global vacuum energy. The energy difference between $U_l$ in the vacuum and its value in the neutral bulk is made up of both the surface dipole energy, $e \Delta\varphi_{surf}$, and the total band bending in the SCR, $e \Delta\varphi_{BB}^{tot}$. The energy difference between $U_l$ in the neutral bulk and $E_F$ is denoted by $W_b$. Note that while the system shown is an $n$-type system with an acceptor state at the surface, the treatment is general.

The treatment proposed is based on the assumption that the only part of the system in \Cref{fgr:theory-main}(a) that requires an explicitly quantum-mechanical, atomistic treatment is the region immediately adjacent to the surface. From a certain position along the surface normal, which we denote by $z_d$, the system can be adequately described using electrostatics. This observation is strongly supported by the excellent agreement between a DFT-based ``doped-atom'' simulation of a heavily-doped $p$-$n$ junction \cite{Sinai2013} and the textbook effective-mass model \cite{Sze2006}, and is in complete agreement with the above-described ``multi-scale VCA'' approach \cite{Richter2013,Moll2013,Xu2013}. We therefore bisect the system of \cref{fgr:theory-main}(a) at $z_d$. This position is chosen such that the short-range atomistic effects of the surface have decayed. The slab remaining to its right will be thin enough to allow its treatment with atomistic detail, e.g., from first principles. The bisection is shown as a vertical dashed line in \cref{fgr:theory-main}(b). Note that because the CBM and VBM always follow $U_l$, we have omitted them in the diagram of \cref{fgr:theory-main}(b) and rescaled it relative to \cref{fgr:theory-main}(a). The potential drop occurring across the left-hand-side semi-infinite bulk is denoted by $\Delta\varphi_b$, with the remainder $\Delta\varphi_{slab}$ occurring across the right-hand-side slab. The proper alignment of the different energy levels in the two parts of the calculation is guaranteed by their common value of $U_l$ at the partition point $z_d$, which we denote by $U_d \equiv U_l(z_d)$.

\subsection*{``Contracting'' the SCR to a charged sheet in a thin slab system}
A correct description of the slab on the right-hand side of the bisection requires not only a quantum-mechanical treatment, but also the explicit inclusion of the effect of the left side of the system. Because the plane of bisection at $z_d$ goes through the SCR, it divides the system, which is overall charge neutral, into two oppositely charged parts. $\Delta Q$ denotes the net charge per unit area of the right-hand-side slab. In the $n$-type case shown in \cref{fgr:theory-main}(b), electronic charge will be transferred to the surface and the sign of $\Delta Q$ will be negative. In a $p$-type case, $\Delta Q$ would be positive. Due to our choice of $z_d$, the only way in which the semi-infinite bulk on the left-hand side affects the slab on the right is through the charge transfer of $\Delta Q$, and the electrostatic field that results from this.

To include these effects, we perform the following gedanken experiment: Beginning from \cref{fgr:theory-main}(b) and given a certain $\Delta Q$, we ``move'' the left-hand part leftward, as shown in \cref{fgr:theory-main}(c), creating a parallel capacitor of charge per unit area $\Delta Q$. Correspondingly, an electric field arises in the vacuum ``layer'' introduced between the two parts of the system. Due to Gauss' law, this field is independent of the width of the vacuum layer and of the particular distribution of the net charge on either side of it. Therefore, the semi-infinite bulk may be ``contracted'' to a uniform, fixed charged sheet of vanishing width, with charge per unit area $Q_{sheet} = -\Delta Q$, as shown in \cref{fgr:theory-main}(d). In this way, the effects of the SCR in the semi-infinite bulk can now be included within an atomistic simulation of the slab. The excess charge $\Delta Q$ is allowed to distribute freely, and is exactly compensated by the fixed sheet of charge $-\Delta Q$, so that the system is overall charge neutral.

Since the effect of the SCR is entirely included via the charged sheet, the key parameter needed for an accurate description of the slab system is the correct, equilibrium value of $\Delta Q$, the charge transfer from the semi-infinite bulk on the left to the slab on the right. To find this, we proceed with the following self-consistent scheme. We begin by making a reasonable guess for the amount of charge transferred \textit{to} the right-hand-side slab, which we denote by $\Delta Q_S$. Using our charged-sheet technique, we calculate the atomistic slab corresponding to the right-hand side of the system, for this $\Delta Q_S$, and denote the Fermi level we obtain from the calculation by $E_F^0$. As shown next, we then use these results to determine the value of the common energy $U_d$. Assuming Fermi level equilibration between the slab and the bulk (as shown in \cref{fgr:theory-main}(b)), we calculate the band-bending in the left-hand side, $\Delta \varphi_b$. From this we determine the corresponding amount of charge transferred \textit{from} the left-hand-side bulk, which we denote by $\Delta Q_B$. The scheme then proceeds until self-consistency between $\Delta Q_S$ and $\Delta Q_B$ is reached. We now explain how this to achieve this in practice.

\subsection*{Finding the common energy $U_d$}
In order to find $U_d$, let us examine the $U_l$ curve in the slab system shown in \cref{fgr:theory-main}(d). The idealized $U_l$, denoted by $U_i(z)$ and marked by the blue curve in \cref{fgr:theory-main}(d), includes only the changes due to the SCR contraction: The semi-infinite bulk $U_l$ profile from \cref{fgr:theory-main}(b) has been replaced by a linear potential drop in the vacuum layer left of the slab, beginning at the charged sheet and ending at the point $(z_d,U_d)$. At $z_d$, $U_i(z)$ changes slope abruptly due to the change in the dielectric constant, $\varepsilon$, from vacuum to dielectric, providing a straightforward way to identify $(z_d,U_d)$. However, treating the slab with electronic structure theory causes the electrostatic profile to differ from this idealized $U_i(z)$. In practice, we extract $U_l$ from our atomistic calculation and denote the result by $U_a(z)$\label{pp:Uaz}. This is achieved by averaging out microscopic variations in the electrostatic energy $U(\vec{r})$, which is an output of the simulation: First, $U(\vec{r})$ is plane-averaged perpendicular to the surface normal throughout the slab supercell, yielding a plane-averaged electrostatic energy $\bar{U}(z)$. Then, an additional out-of-plane averaging is performed over an interval corresponding to a unit cell of the bulk material. This results in the macroscopic electrostatic energy \cite{Baldereschi1988}, which corresponds to $U_l$. A typical result for $U_a(z)$ from an atomistic simulation is shown schematically as the red curve in \cref{fgr:theory-main}(d) (slightly offset from $U_i(z)$ to make it easier to differentiate them in the intervals where they coincide). Comparison of $U_a(z)$ to $U_i(z)$ highlights important differences between them:

\begin{enumerate}
  \item The distinction between between $\Delta\varphi_{surf}$ and $\Delta\varphi_{slab}$ is no longer clear; this is in keeping with the blurring of the distinction between band-bending and surface dipole at microscopic distances from the surface \cite{Yaffe2013}. We denote their sum by $\Delta \varphi_S^{tot}$.
  \item The introduction of an interface between the slab and the left-hand vacuum (towards the charged sheet) causes a spurious potential step, $\Delta\varphi_{v}$, to arise, due to the dipole at this ``virtual'' surface. Furthermore, charge reorganization near the virtual surface now explicitly accounts for the change in the slope of $U_a(z)$, which in $U_i(z)$ is a result of the change in dielectric constant $\varepsilon$.
  \item Based on the curve for $U_i(z)$, $z_d$ may be viewed as the ``formal left edge'' of the slab, i.e. the position at which the slab ``ends'' and the (spurious) left-hand vacuum ``begins''. However, $U_a(z)$ does not generally go through the point $(z_d, U_d)$.
\end{enumerate}

Note that as a practical consideration, the virtual surface must be fully passivated to prevent the creation of spurious gap-states (e.g. by capping danging bonds with hydrogens or pseudo-hydrogens \cite{Huang2005}).

As just emphasized, $U_d$ cannot be simply read off of $U_a(z)$. However, if we were able to reconstruct $U_i(z)$ from our available $U_a(z)$ data, we would be able to identify the point $(z_d, U_d)$ by the change of slope as mentioned above. In other words, if we reconstruct the curves corresponding to $U_i(z)$ in the slab and $U_i(z)$ in the vacuum to the left of the slab, respectively, then $(z_d, U_d)$ will be their intersection.

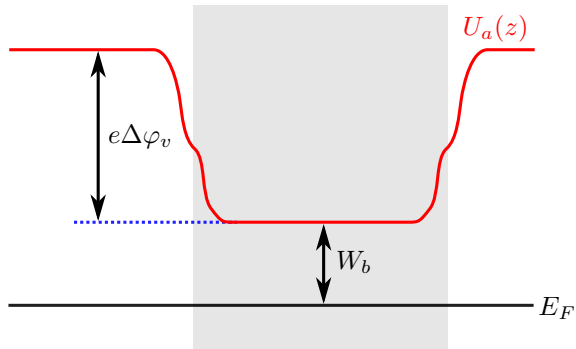
\begin{figure}[htb!]
  \hfill
\begingroup%
  \setlength{\unitlength}{0.5\textwidth}%
  \begin{picture}(1,0.62515289)%
    \put(0,0){\includegraphics[width=\unitlength]{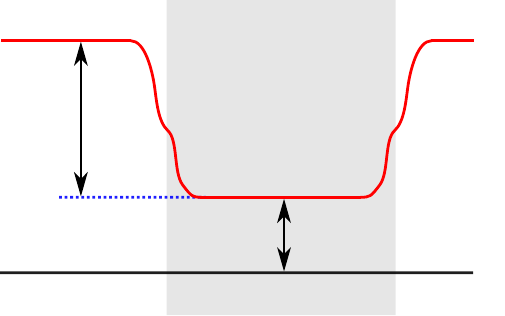}}%
    \put(0.94715663,0.06672022){\color[rgb]{0,0,0}\makebox(0,0)[lb]{\smash{$E_F$}}}%
    \put(0.17802355,0.37740169){\color[rgb]{0,0,0}\makebox(0,0)[lb]{\smash{$e\Delta\varphi_v$}}}%
    \put(0.5852746,0.14955965){\color[rgb]{0,0,0}\makebox(0,0)[lb]{\smash{$W_b$}}}%
    \put(0.81200549,0.57178025){\color[rgb]{1,0,0}\makebox(0,0)[lb]{\smash{$U_a(z)$}}}%
  \end{picture}%
\endgroup%
  \hspace*{\fill}
  \caption{A slab system similar to the system in \cref{fgr:theory-main}(d), but fully passivated on both sides. Key values are shown (see text for details).}
  \label{fgr:theory-FP}
\end{figure}

In the slab, it is possible to reconstruct $U_i(z)$ using the fact that the slab is constructed of ``doped'' atoms with fractional $Z$ matching the nominal bulk doping, while making the slab thick enough for its central region to be bulk-like. The idealized $U_i(z)$ to the right of $z_d$ can then be extrapolated from $U_a(z)$ in a central bulk-like region, where the two are expected to match (the interval where the red and blue curves overlap in \cref{fgr:theory-main}(d)). In order to reconstruct $U_i(z)$ in the vacuum, we must first extrapolate a line from $U_a(z)$ in the vacuum between the slab's virtual face and the charged sheet. We denote this line by $U^{vac}(z)$ (indicated by a dashed red line in \cref{fgr:theory-main}(d)). However, in order to bring it into alignment with the idealized $U_i(z)$ in the vacuum, $U^{vac}(z)$ must be corrected for the potential drop on the virtual side, $\Delta\varphi_{v}$\bibnote{
$z_d$ is the intersection of two curves that together comprise $U_i(z)$: A straight line in the vacuum left of the slab and a curved line in the slab. The slope of the vacuum line corresponds to the sheet charge, and therefore one point on that line is sufficient to reconstruct it. We approximate the curved line by a parabola, allowing us to reconstruct it using three points on $U_a(z)$, taken from the region where it is reasonable to assume that it matches $U_i(z),$ i.e., in a bulk-like interval in the slab.}.
This latter parameter can be obtained from a calculation of a slab system as similar as possible to the system investigated, but fully passivated on both slab faces (excluding of course any charged sheet). This is shown in \cref{fgr:theory-FP}. We make the assumption that the change in $\Delta\varphi_{v}$ due to the presence of a field is negligible. This is a reasonable assumption because the macroscopic field is small relative to microscopic, atomistic fields. Furthermore, it is borne out by our results (see below). We then find the point $(z_d,U_d)$ as the intersection of $U^{vac}(z) - e\Delta\varphi_v$ with the extrapolation of $U_a(z)$ from the bulk-like region of the slab.

We have found that this intersection-of-curves procedure is robust even if bulk-like conditions are not strictly achieved within the slab (which we deduce by observing that the slope of $U_a(z)$, extrapolated from within the slab, differs from its slope in vacuum divided by the bulk dielectric constant). This typically creates only a minor deviation in the estimate of the slab's ``formal edge'', i.e., of $z_d$. In our various numerical tests, described in detail below, we found that even an error as large as 2 \A\ in $z_d$ leads to no more, and usually much less, than a $\sim$5 meV error in the converged position of the slab's Fermi level, which is the most sensitive slab parameter examined.

\subsection*{Finding the net charge removed from the bulk, $\Delta Q_B$}
Having determined $U_d$, we use this value, along with $E_F^0$ - the Fermi level from the atomistic calculation - to find $\Delta\varphi_b$ based on \cref{fgr:theory-main}(b):
\begin{equation}
  \label{eq:bulk-dphi_b-of-Ef}
  \Delta\varphi_b = \frac{1}{e} (U_d - W_b - E_F^0)
\end{equation}
The energy difference between the Fermi level and the local vacuum level in the neutral bulk, $W_b$, which is needed for using the above equation in practice, is also easily extracted from the fully-passivated system in \cref{fgr:theory-FP}. As we have shown previously \cite{Sinai2013}, in a fully passivated slab the ``doped'' atoms (which reflect the bulk nominal doping) determine $E_F$ uniquely.

Using the value of $\Delta\varphi_b$ we obtained from \cref{eq:bulk-dphi_b-of-Ef}, we now turn to calculate the amount of charge \textit{removed} from the left-hand side, $\Delta Q_B$, which, unless self-consistency has already been reached, will generally differ from $\Delta Q_S$ (note that because $\Delta Q_S$ and $\Delta Q_B$ refer to charge inserted and removed, respectively, they are of the same sign). We treat the left-hand side of the system using the effective-mass approximation, which provides a good electrostatic description of the semi-infinite bulk. Specifically, we solve the Poisson-Boltzmann equation for this side \cite{Sze2006}:
\begin{equation}
  \label{eq:P-B}
  \frac{d^2 \left(\Delta \varphi \right)}{dz^2} = -\frac{e}{\varepsilon\varepsilon_0} \left( n_0 - p_0 + p(z) - n(z) \right)
\end{equation}
where $e$ is the electron charge and $\varepsilon_0$ is the vacuum permittivity. $\Delta \varphi$ is the $z$-dependent potential drop, which at $z_d$ attains the value $\Delta \varphi_b$ (see \cref{fgr:theory-main}(b)). $n(z)$ and $p(z)$ are carrier concentrations, given by
\begin{equation}
  \label{eq:P-Bpn}
  \begin{aligned}
  p(z) &= N_V \times F_{1/2}\left(\frac{-e \Delta \varphi(z) - [E_F - E_{VBM}]^{bulk}}{kT}\right)\\
  n(z) &= N_C \times F_{1/2}\left(\frac{e \Delta \varphi(z) + [E_F - E_{VBM}]^{bulk} - E_g}{kT}\right)
  \end{aligned}
\end{equation}
where $F_{1/2}(\eta_F) \equiv \frac{2}{\sqrt{\pi}} \int_0^\infty \frac{\eta^{1/2}}{1+\exp(\eta - \eta_F)}d\eta$ is the Fermi-Dirac integral, $k$ is Boltzmann's constant, and $T$ is the temperature. The position of the Fermi level relative to the VBM in the neutral bulk, $[E_F - E_{VBM}]^{bulk}$, is found from the requirement of charge neutrality in the bulk, based on the doping density (which is known). The expressions for the bulk carrier concentrations, $n_0$ and $p_0$, can be obtained by taking $\Delta \varphi(z) = 0$ in \cref{eq:P-Bpn}. The bulk parameters, namely the effective density of states near the valence and conduction band edges, $N_V$ and $N_C$, respectively, the dielectric constant $\varepsilon$, and the gap energy $E_g$, can, if so desired, be obtained from separate first-principles calculations. Alternatively, they may be taken from reliable experimental data. Note, however, that due to quantum-size effects \cite{Delley1995}, $E_g$ in slab and bulk calculations may differ. Therefore, in order to maintain consistency between both sides of the system, it is preferable to extract also $E_g$ from the fully-passivated system shown in \cref{fgr:theory-FP}.

\Cref{eq:P-B,eq:P-Bpn} can be solved to yield an expression for the total space charge per unit area in the semi-infinite bulk, equivalent to $-\Delta Q_B$, as a function of the bulk band bending $\Delta\varphi_b$. In cases where the doping is not degenerate, an analytical solution may be derived \cite{Sze2006}:
\begin{equation}
  \label{eq:bulk-dQ}
  \Delta Q_B(\Delta \varphi_b) = sign(\Delta\varphi_b) \frac{\sqrt{2}\varepsilon\varepsilon_0 kT}{eL_D} F\left(\beta\Delta\varphi_b, \frac{n_0}{p_0}\right)
\end{equation}
where $\beta = \frac{e}{kT}$, $L_D$ is the extrinsic Debye length for holes,
\begin{equation*}
  L_D \equiv \sqrt{\frac{kT\varepsilon\varepsilon_0}{p_0 e^2}}
\end{equation*}
and
\begin{equation*}
  F\left(\beta \Delta\varphi_b, \frac{n_0}{p_0}\right) = \sqrt{\left[ \exp(-\beta \Delta\varphi_b) + \beta \Delta\varphi_b - 1 \right] + \frac{n_0}{p_0}[exp(\beta \Delta\varphi_b) - \beta \Delta\varphi_b - 1]}
\end{equation*}
This solution is based on the fact that the Fermi-Dirac integral, $F_{1/2}(\eta_F)$ in \cref{eq:P-Bpn}, approaches $\exp(\eta_F)$ in non-degenerate doping conditions. In the degenerate case, \cref{eq:P-B,eq:P-Bpn} must be solved numerically.

If we now insert \cref{eq:bulk-dphi_b-of-Ef} into \cref{eq:bulk-dQ} (or into a numerical solution of \cref{eq:P-B,eq:P-Bpn}), we obtain $\Delta Q_B$, which we then compare to $\Delta Q_S$. If charge conservation has been obtained, i.e., $\Delta Q_S = \Delta Q_B = \Delta Q$, then our entire system is self-consistent. If they differ, a new guess for $\Delta Q_S$ is chosen, and the process is repeated, until self-consistency is reached.

In principle, the guess for $\Delta Q_S$ in each step may be set to, e.g., $\Delta Q_B$ obtained in the previous step. In practice, however, the convergence of the self-consistent cycle can be accelerated by making use of the density of states (DOS) of the slab calculated in each step. The amount of excess charge in the slab as a function of \textit{any} Fermi level position, $\Delta Q_S'$ (where the prime indicates that it is variable), is easily obtained by energy integration of the slab's DOS. The amount of excess charge from the bulk as a function of the Fermi level position, $\Delta Q_B'$, is already known from \cref{eq:bulk-dphi_b-of-Ef,eq:P-B,eq:P-Bpn}. We can then find a Fermi level position for which $\Delta Q_S' = \Delta Q_B'$ and use this excess charge value as a guess for $\Delta Q_S$ in the next step. While the DOS curve can and sometimes does itself depend on the sheet charge (see numerical examples below), this dependence is typically weak and therefore convergence is rapid. We further note that to ensure consistency with the effective-mass treatment of the left-hand-side bulk, Fermi-Dirac statistics at the correct temperature must be enforced in the slab. In the direct iteration scheme, this must be done within the electronic structure calculation itself, because the Fermi level position obtained from the slab calculation is used explicitly (in \cref{eq:bulk-dphi_b-of-Ef}). If instead one integrates over the resulting DOS (using Fermi-Dirac statistics), any desired eigenstate occupation ``smearing'' scheme, commonly used to facilitate numerical convergence \cite{Verstraete2001}, can be used within the atomistic calculation.

\subsection*{Correcting the total energy}
The achievement of self-consistency between $\Delta Q_S$ and $\Delta Q_B$ ensures that the effects of the SCR in the semi-infinite bulk are correctly included in the atomistic calculation of the slab, and this will also be reflected in the total energy of the system. However, because the SCR is not explicitly included in the atomistic calculation, the raw total energy result coming out of the atomistic calculation, $E_a^{tot}$, must be corrected. We proceed to derive the total energy correction required, beginning from the bisected system in \cref{fgr:theory-main}(b). The total energy of this system, $E^{tot}$, can be divided into two contributions, namely, the total energy of the semi-infinite bulk ($z < z_d$) and that of the slab ($z > z_d$). From our basic assumption, namely, that an electrostatic description is sufficient at $z < z_d$, it follows that $E^{tot}(z < z_d)$ can be found by integrating the square of the electrostatic field, $\mathbf{F}(z)$, along the surface normal. Therefore,
\begin{equation}
\label{eq:Etot-scr}
  E^{tot} = \frac{1}{2} \varepsilon \varepsilon_0 \int_{-\infty}^{z_d} \mathbf{F}^2(z) dz + E_s^{tot}
\end{equation}
where $E_s^{tot}$ denotes the total energy of the $z > z_d$ slab. While $E_s^{tot}$ must be derived from the atomistic treatment, it does not equal the total energy obtained from it directly, denoted above by $E_a^{tot}$. The latter includes, in addition to the slab itself, the energy of the thin sheet of charge $\Delta Q$, placed at a certain position, $z_{sheet}$, left of $z_d$. Furthermore, charge rearrangement takes place owing to the creation of the virtual surface (this is also associated with the appearance of the spurious potential drop, $\Delta\varphi_v$ - recall \cref{fgr:theory-main}(d)). Thus,
\begin{equation}
\label{eq:Etot-a}
  E_a^{tot} = E_s^{tot} + \frac{1}{2} \varepsilon_0 \int_{z_{sheet}}^{z_d} \mathbf{F}_c^2 dz + \Delta E_v^{tot}
\end{equation}
where $\mathbf{F}_c$ is the (constant) field in the vacuum layer between $z_{sheet}$ and $z_d$, and $\Delta E_v^{tot}$ denotes the energy required to form the virtual surface. Combining \cref{eq:Etot-scr,eq:Etot-a}, we obtain:
\begin{equation}
\label{eq:Etot}
  \begin{aligned}
  E^{tot} &= \frac{1}{2} \varepsilon \varepsilon_0 \int_{-\infty}^{z_d} \mathbf{F}^2(z) dz + E_a^{tot} - \frac{1}{2} \varepsilon_0 \int_{z_{sheet}}^{z_d} \mathbf{F}_c^2 dz - \Delta E_v^{tot}\\
  &= E_a^{tot} + E_{SCR}^{tot} - E_c^{tot} - \Delta E_v^{tot}
  \end{aligned}
\end{equation}
The first term on the right-hand side of this equation, $E_a^{tot}$, is obtained directly from the atomistic calculation. The second term, $E_{SCR}^{tot} \equiv \frac{1}{2} \varepsilon \varepsilon_0 \int_{-\infty}^{z_d} \mathbf{F}^2(z) dz$, can be calculated once $\mathbf{F}(z)$ in the SCR is obtained from a numerical solution of \cref{eq:P-B,eq:P-Bpn}. The third term, $E_c^{tot}$, can be calculated analytically from the known values of $\Delta Q$, $z_d$, and $z_{sheet}$ as:
\begin{equation*}
  E_c^{tot} \equiv \frac{1}{2} \varepsilon_0 \int_{z_{sheet}}^{z_d} \mathbf{F}_c^2 dz = \frac{1}{2} \varepsilon_0 (z_d - z_{sheet}) \mathbf{F}_c^2 = \frac{z_d - z_{sheet}}{2 \varepsilon_0} \Delta Q^2
\end{equation*}
The final term in \cref{eq:Etot}, $\Delta E_v^{tot}$, is generally unknown. However, in keeping with our previous assumption in the context of the virtual surface, namely, that the macroscopic field is small relative to microscopic/atomistic fields, we may assume that $\Delta E_v^{tot}$ is constant with respect to reasonable values of $\Delta Q$. As a result, it will cancel out when comparing different surfaces of the same semiconductor. Therefore, the correction we must apply to $E_a^{tot}$ to regain the correct total energy, up to an uninteresting constant, is 
\begin{equation}
\label{eq:dEtot}
  \Delta E^{tot} \equiv E_{SCR}^{tot} - E_c^{tot}
\end{equation}

In order for \cref{eq:dEtot} to apply, care must be taken to ascertain that the surfaces compared possess the same bulk properties and the same virtual-side passivation.

\subsection*{Implementation}
We have implemented CREST as a wrapper around an atomistic electronic structure program. To be compatible with the wrapper, the electronic structure program must allow for the inclusion of a fixed sheet of charge, $Q_{sheet}$, as well as for the use of pseudoatoms with a fractional $Z$ that serve as the ``doped'' atoms. The atomistic program is otherwise arbitrary in both the type of electronic structure approach used (e.g., empirical pseudopotentials, tight binding, density functional theory, etc.) and in other implementation details (e.g., basis set, treatment of core electrons, etc.). In practice, communication between the wrapper and the atomistic program occurs only once in each self-consistent step, and comprises running the program with a sheet of charge $Q_{sheet}^{in}$ (and added free charge of $-Q_{sheet}^{in}$), and thereafter extracting $U_a(z)$ and the slab DOS. This is done through an interface script, which must be implemented for each specific atomistic electronic structure program.

The atomistic slab is constructed by the user such that one side of the slab is the surface of interest and the other is a passivated, virtual surface, with ``doped'' semiconductor atoms within the slab. The charged sheet is placed to the left of the virtual surface, far enough from it such that no charge density is found in its vicinity (the exact position of the sheet does not affect the physics, as long as periodic boundary conditions parallel to the surface are set). The user must also provide $W_b$ and $\Delta \varphi_v$ obtained from a calculation of the fully passivated slab (see \cref{fgr:theory-FP}), which only needs to be performed once for the entire self-consistent cycle.

\section{Tight-binding electronic structure calculations}
\label{sec:TB}
As a first step, we wish to ascertain that CREST indeed reproduces accurately results from calculations in which the SCR is entirely included within the calculation cell, across a wide range of scenarios. Because the technique outlined in the theory section is in principle compatible with any atomistic electronic structure approach chosen, it is most easily tested with a computationally inexpensive one. In this section we therefore use CREST as a wrapper around a simple, 1-D tight-binding (TB) code. It has the advantage of being computationally minimal while still reliably capturing the long-range electrostatic behavior. As a result, we are able to explicitly calculate systems whose SCR is much wider, relative to the distances between atoms, than would be possible within a first-principles (e.g., DFT) treatment. These results then serve as a reference point for calculations of slabs which are thin relative to the SCR, both without and with CREST.

\subsection*{Description of the TB model system}
The Hamiltonian of a model system of N sites is described by:
\begin{equation}
\label{eq:H-all}
\mtx{H} = \mtx{H}_{TB} + \mtx{U}_{ES}
\end{equation}
The first part is the TB Hamiltonian. In the TB system, each ``atom'' (TB site) possesses two single-particle levels with different on-site energies, denoted by $E^I_j$ and $E^{II}_j$. Atoms interact only with their two nearest neighbors, via the interaction energies $S^I_{j,j+1}$, $S^{II}_{j,j+1}$ and $S^{I,II}_{j,j+1}$. The former two denote interaction between similar states on neighboring sites, while the latter indicates cross-interaction of different states on neighboring sites. There is no interaction between different states on the same site. This leads to the following form of TB Hamiltonian:
\begin{multline}
\label{eq:H-TB}
\mtx{H}_{TB} = 
\sum_{j = 1}^N \sum_{k = I,II} E^k_j\ket{j,k}\bra{j,k} + \\
\sum_{j = 1}^{N-1} \left[ \sum_{k = I,II} S^k_{j,j+1} \left( \ket{j,k}\bra{j+1,k} + \ket{j+1,k}\bra{j,k} \right) + \right. \\
\left. S^{I,II}_{j,j+1} \sum_{k \neq k'} \left( \ket{j,k}\bra{j+1,k'} + \ket{j+1,k}\bra{j,k'} \right) \right]
\end{multline}
which is given in explicit matrix form by:
\begin{equation*}
\mtx{H}_{TB} = 
\left[
  \begin{array}{cccccc}
    \mtx{E}_1 & \mtx{S}_{12} & \mtx{0} & \cdots & \mtx{0} & \mtx{0}\\
    \mtx{S}_{12} & \mtx{E}_2 & \mtx{S}_{23} & \cdots & \mtx{0} & \mtx{0}\\
    \mtx{0} & \mtx{S}_{23} & \mtx{E}_3 & \cdots & \mtx{0} & \mtx{0}\\
    \vdots & \vdots & \vdots & \ddots & \vdots & \vdots\\
    \mtx{0} & \mtx{0} & \mtx{0} & \cdots & \mtx{E}_{N-1} & \mtx{S}_{N,N-1}\\
    \mtx{0} & \mtx{0} & \mtx{0} & \cdots & \mtx{S}_{N,N-1} & \mtx{E}_N\\
  \end{array}
\right]
\end{equation*}
where $\mtx{0}$ is a 2x2 zero matrix and $\mtx{E}_j$, $\mtx{S}_{j,j+1}$ are the 2x2 matrices:
\begin{equation*}
\mtx{E}_j = 
\left[
  \begin{array}{cc}
    E^I_j & 0\\
    0 & E^{II}_j
  \end{array}
\right]
\qquad \qquad
\mtx{S}_{j,j+1} = 
\left[
  \begin{array}{cc}
    S^I_{j,j+1} & S^{I,II}_{j,j+1}\\
    S^{I,II}_{j,j+1} & S^{II}_{j,j+1}
  \end{array}
\right]
\end{equation*}

In general, $\mtx{E}_j$ are identical for all $j$ belonging to ``bulk atom'' sites, as are all $\mtx{S}_{j,j+1}$. We denote the bulk-atom lower and higher energies by $E^I_b$ and $E^{II}_b$, and the inter-level interactions by $S^I_b$, $S^{II}_b$ and $S^{I,II}_b$, respectively. If the on-site energies are far enough apart relative to the magnitude of the inter-level interactions and $S^I_b \approx S^{II}_b \approx S^{I,II}_b$, then a semiconductor results, with a forbidden gap of roughly $E_g \cong E^{II}_b - E^I_b - 2(|S^I_b| + |S^{II}_b|)$ (slightly increased by the cross-level $S^{I,II}_{j,j+1}$ interactions). We note that this system bears some similarity to the alternating-site TB model semiconductor presented in Ref. \cite{Landau2009}. In our TB model, we illustratively used $E^I_b = -5 \text{ eV}$, $E^{II}_b = -3 \text{ eV}$, $S^I_b = -0.2 \text{ eV}$, $S^{II}_b = -0.3 \text{ eV}$ and $S^{I,II}_b = -0.1 \text{ eV}$. These result in a bulk forbidden gap of about $E_g \cong 1 \text{ eV}$.

Vacuum is represented by ``empty'' sites, i.e. sites with very high on-site energies which do not interact with their neighbors ($\mtx{S}_{j,j+1} = \mtx{0}$). We use 10 vacuum sites on either end of the chain in all systems. A defect with a gap state is represented by a single, non-interacting site with one energy level within the TB semiconductor gap. Specifically, we place a defect site between a bulk atom site and a vacuum site to mimic a surface defect.

The second part of our model Hamiltonian (right side of \cref{eq:H-all}) is an electrostatic term which shifts the on-site energies $E^I_j$ and $E^{II}_j$, and has the form
\begin{equation}
\label{eq:H-U}
\mtx{U}_{ES} = 
\sum_{j = 1}^N U_j \sum_{k = I,II} \ket{j,k}\bra{j,k}
\end{equation}
The values of $U_j$ are determined as follows. We allow each energy level to be occupied by one electron only (no spin degeneracy). Each undoped bulk-atom (site) is attributed a ``nuclear'' (positive) charge of 1, and contributes 1 electron to the system. Vacuum sites are attributed no nuclear charge and contribute no electrons. The correct charging behavior of defect gap states is achieved as follows: If an electron acceptor defect is desired, the non-gap energy level is positioned well within or below the semiconductor VB, such that it is always full. This way, if the gap level is empty the acceptor is neutral and if it is filled the acceptor is negatively charged, as appropriate. Similarly, for a donor the non-gap energy level is placed well within or above the CB, such that it is always empty, making the donor positively charged or neutral. 

Once the system's eigenstates are found, its $E_F$ is determined by occupying them, using the Fermi-Dirac distribution at room temperature, up to charge neutrality. The negative charge at each site is determined by projecting the system's (delocalized) occupied eigenstates onto each site. If the fixed, positive nuclear charges are added to this electronic charge profile, a net charge profile as a function of site position is obtained. This site-based charge profile is transformed into a $z$-dependent net charge-per-area profile, $\rho(z)$, by attributing an arbitrary uniform distance between neighboring sites and a lateral area to the chain (in our model system, 2 \A\ and 4 \A$^2$, respectively). We calculate the electrostatic potential of the system, $\varphi(z)$, from $\rho(z)$ using the 1-D Poisson equation, $\frac{d^2}{dz^2} \varphi(z) = -\frac{1}{\varepsilon_0} \rho(z)$. The local vacuum level $U_l(z)$ is taken to be equal to $e\varphi(z)$, and taking this at the site positions yields the $U_j$'s of \cref{eq:H-U}. The system is recalculated until self-consistency, i.e. until solving the Hamiltonian of the system at a given $\mtx{U}_{ES}$ yields a charge profile $\rho(z)$ which, in turn, results in that same $\mtx{U}_{ES}$.

In this model system, the value of $N_{ss}$, the surface-state density, can be changed by attributing a weight to the projection of the system eigenstates on to a surface-defect site. This weight thus modifies the maximum possible occupation of the defect levels. In addition, it modifies the nuclear and electronic charges attributed to the site. The charged sheet required for CREST calculations is represented simply by a site similar to the vacuum sites described above, but with a nonzero fixed (negative or positive) charge. Because the on-site energies are very high, the associated electronic charge is automatically diverted to the slab itself, where it distributes freely. Finally, bulk doping is handled using fractional modifications to the nuclear charge of the atoms and corresponding increase/decrease of electronic charge (i.e., by using the above-described concept of ``doped'' atoms). This affects the Hamiltonian via $\mtx{U}_{ES}$. A connection is made between the ``doped atom'' modifications and the volumetric doping rate $N_{dop}$, as well as other quantities needed for the effective-mass treatment, via our selected distance between sites and chain lateral area. As expected, minute modifications (moderate doping) cause negligible alterations in the TB eigenvalues, while $E_F$ shifts within the gap.

In order to obtain the bulk parameters $N_V$, $N_C$ and $\varepsilon$ for this fictitious system, we first calculated explicitly the semiconductor bulk, as well as symmetric $p$-$n$ junctions, at various doping densities. We then compared the results from these calculations with effective-mass calculations, and derived the bulk parameter values which yielded the best fit. The value we obtained for $\varepsilon$ was about 1.5, indicating relatively low screening.

\subsection*{SCR confinement and CREST results in simulations of a typical system}
\begin{figure}[htb!]
\centering
  \begin{tabular}{ c m{0.6\textwidth} }
    (a) &
    \pstool![width=0.6\textwidth]{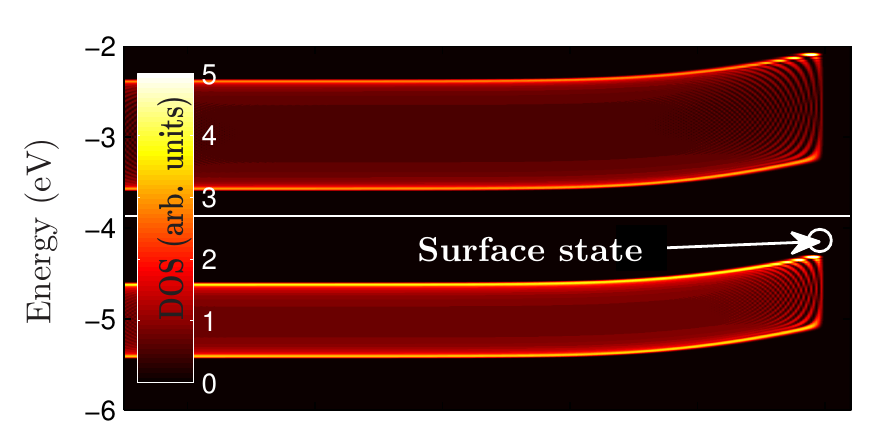}{
      \psfrag{y1}[b][b]{\setlength{\tabcolsep}{0pt}\begin{tabular}{c}
        Energy (eV)
      \end{tabular}}
      \psfrag{y2}[c][c]{\setlength{\tabcolsep}{0pt}\begin{tabular}{c}
        ~\scalebox{0.75}[1.0]{\textbf{DOS (arb. units)}}
      \end{tabular}}
      \psfrag{lb1}[r][r]{\setlength{\tabcolsep}{0pt}\begin{tabular}{c}
        \textcolor[rgb]{1,1,1}{\textbf{Surface state}}~
      \end{tabular}}
    }\\
    (b) &
    \pstool![width=0.6\textwidth]{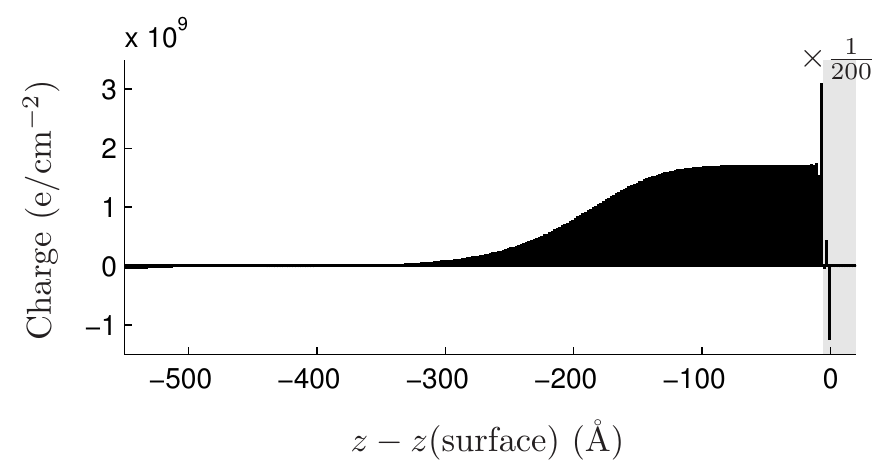}{
      \psfrag{x1}[t][t]{\setlength{\tabcolsep}{0pt}\begin{tabular}{c}
        $z - z(\text{surface})$ (\A)
      \end{tabular}}
      \psfrag{y1}[b][b]{\setlength{\tabcolsep}{0pt}\begin{tabular}{c}
        Charge (e/cm$^{-2}$)
      \end{tabular}}
      \psfrag{lb1}[b][b]{\setlength{\tabcolsep}{0pt}\begin{tabular}{c}
        $\times \frac{1}{200}$
      \end{tabular}}
    }\\
  \end{tabular}
  \caption{Illustrative example of an $n$-doped system, calculated using our 1-D TB code ($N_{dop} = 1.25 \times 10^{17} \text{cm}^{-3}$, $N_{ss} = 2.5 \times 10^{11} \text{cm}^{-2}$) with a slab of 300 sites (600 \A), thick enough to include the full SCR: (a) The site-projected DOS (PDOS) showing the valence band (VB) and the conduction band (CB) with the gap between them. $E_F$ is indicated by the horizontal white line. The bands are bent by the field in the SCR. The acceptor surface state resides on the right side of the slab. As it is invisible relative to the color scale, its position is marked on the figure. (b) Column plot showing the net charge per area associated with each ``atom'' in the TB treatment. In the shaded region on the right (adjacent the surface), charge magnitudes are scaled down by a factor of 200.}
  \label{fgr:TB-example-fullSCR}
\end{figure}

We select a typical system comprising an $n$-type material with $N_{dop} = 1.25 \times 10^{17} \text{cm}^{-3}$ and a deep acceptor surface defect with $N_{ss} = 2.5 \times 10^{11} \text{cm}^{-2}$, in order to illustrate typical trends in systems calculated with our TB code. Some results from the ``full-SCR'' calculation of this system, i.e., with a slab which is sufficiently thick to include the full SCR, are shown in \cref{fgr:TB-example-fullSCR}.

\begin{figure}[htb!]
\centering
  \begin{tabular}{ l l }
    (a) & (b)\\
    \pstool![width=0.45\textwidth]{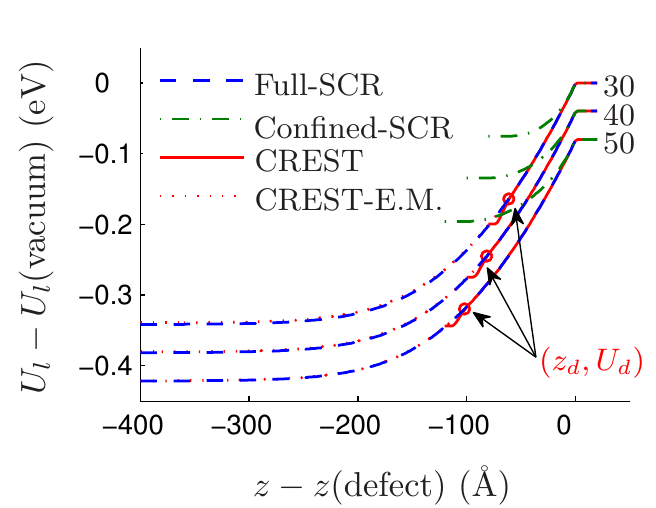}{
      \psfrag{x1}[t][t]{\setlength{\tabcolsep}{0pt}\begin{tabular}{c}
        $z - z(\text{defect})$ (\A)
      \end{tabular}}
      \psfrag{y1}[b][b]{\setlength{\tabcolsep}{0pt}\begin{tabular}{c}
        $U_l - U_l(\text{vacuum})$ (eV)
      \end{tabular}}
      \psfrag{blue}[l][l]{\setlength{\tabcolsep}{0pt}\begin{tabular}{c}
        \small{Full-SCR}
      \end{tabular}}
      \psfrag{green}[l][l]{\setlength{\tabcolsep}{0pt}\begin{tabular}{c}
        \small{Confined-SCR}
      \end{tabular}}
      \psfrag{red}[l][l]{\setlength{\tabcolsep}{0pt}\begin{tabular}{c}
        \small{CREST}
      \end{tabular}}
      \psfrag{reddotted}[l][l]{\setlength{\tabcolsep}{0pt}\begin{tabular}{c}
        \small{CREST-E.M.}
      \end{tabular}}
      \psfrag{lb1}[l][l]{\setlength{\tabcolsep}{0pt}\begin{tabular}{c}
        \small{30}
      \end{tabular}}
      \psfrag{lb2}[l][l]{\setlength{\tabcolsep}{0pt}\begin{tabular}{c}
        \small{40}
      \end{tabular}}
      \psfrag{lb3}[l][l]{\setlength{\tabcolsep}{0pt}\begin{tabular}{c}
        \small{50}
      \end{tabular}}
      \psfrag{lb4}[l][l]{\setlength{\tabcolsep}{0pt}\begin{tabular}{c}
        \small{\textcolor[rgb]{1,0,0}{$(z_d,U_d)$}}
      \end{tabular}}
    }&
    \pstool![width=0.45\textwidth]{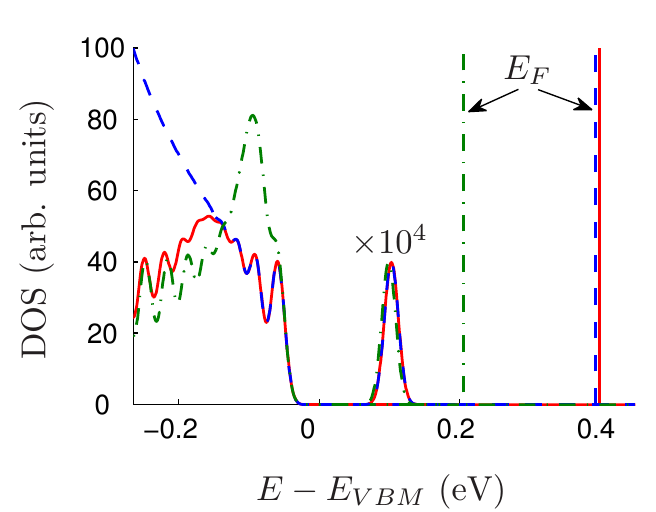}{
      \psfrag{x1}[t][t]{\setlength{\tabcolsep}{0pt}\begin{tabular}{c}
        $E - E_{VBM}$ (eV)
      \end{tabular}}
      \psfrag{y1}[b][b]{\setlength{\tabcolsep}{0pt}\begin{tabular}{c}
        DOS (arb. units)
      \end{tabular}}
      \psfrag{lb1}[b][b]{\setlength{\tabcolsep}{0pt}\begin{tabular}{c}
        $\times 10^4$
      \end{tabular}}
      \psfrag{lb2}[b][b]{\setlength{\tabcolsep}{0pt}\begin{tabular}{c}
        $E_F$
      \end{tabular}}
    }\\
  \end{tabular}
  \caption{Comparison of ``full-SCR'' (blue dashed line), ``confined-SCR'' (green dash-dotted line) and CREST-corrected (red solid line) calculations of the system of \cref{fgr:TB-example-fullSCR}. (a) Electrostatic energy $U_l$ computed from the full-SCR system and from thin slabs containing 30, 40 or 50 TB sites (denoted by ``30'', ``40'' or ``50'' respectively). The sets of curves are offset 0.04 eV from each other for clarity, with the same reference full-SCR result appearing in each set. Red round markers indicate the positions of $(z_d, U_d)$ in each case, with the red dotted lines (denoted by CREST-E.M. in the legend) extending to the left of $z_d$ showing $U_l$ in the effective-mass bulk, obtained by solving \cref{eq:P-B,eq:P-Bpn} numerically within CREST. (b) Total DOS in the vicinity of the VBM. Results from the full system and from the 30-site slab are shown. Vertical lines indicate the Fermi level position in each calculation. The defect level inside the gap is enhanced by a factor of $10^4$. DOS curves are broadened by convolution with a gaussian function with a standard deviation of 0.01 eV.}
  \label{fgr:TB-example-compare}
\end{figure}

The band bending associated with the SCR is clearly visible in the site-projected DOS (PDOS), shown in \cref{fgr:TB-example-fullSCR}(a). The corresponding net charge profile is shown in \cref{fgr:TB-example-fullSCR}(b); a depletion region is discernible as a region of net positive charge, tapering off towards the back (left side) of the slab. These results demonstrate that our TB code reproduces the typical electrostatic features of doped surfaces. The electrostatic energy $U_l$ of this system (which follows the band bending shown in \cref{fgr:TB-example-fullSCR}(a)) and its total DOS are shown in \cref{fgr:TB-example-compare} (dashed blue curves) as our reference data for thin-slab treatments.

To investigate the effect of using an insufficiently thick slab for modeling the surface in \cref{fgr:TB-example-fullSCR}, we calculate it with slabs whose thicknesses are 50, 40 and 30 sites, respectively, i.e., from one sixth to one tenth of the full-SCR system, both without and with the use of CREST, and compare the results in \cref{fgr:TB-example-compare}(a). In the former ``confined-SCR'' case, for all thicknesses the band-bending profile is visibly distorted relative to the full-SCR curve, ``straightening out'' after some fraction of the overall potential energy difference in the full-SCR system of 340 meV. Naturally, the greatest error occurs in the thinnest slab, where the energy difference is a mere 70 meV, one fifth of the full-SCR value. This underestimation is less severe when using thicker slabs, but is naturally greater at lower doping rates we examined (not shown for brevity). An examination of the DOS of the confined-SCR slab in the 30-site case (\cref{fgr:TB-example-compare}(b)) shows that the distortion is associated with a qualitatively erroneous description of the top of the valence band, as well as with an error in the value of $E_F$, which is 0.2 eV too low. It should be noted that because the calculation of the surface work function from first principles depends on the position of $E_F$ and on the electrostatic energy \cite{fall1999}, we also expect this value to be erroneous. In addition, there is a small error ($\sim$5 meV) in the surface state position in the gap. Generally, the surface state position is shifted within the bulk gap by the SCR-related field (as also observed by Kempisty and co-workers \cite{Kempisty2011, Kempisty2011e}). But in the confined-SCR system, this field is too small, leading to the error observed. Similar errors occur in the thicker slabs (not shown for brevity).

As evident from the solid red curves in \cref{fgr:TB-example-compare}, the application of CREST brings the thin slab simulation results into near-perfect agreement with the more expensive thick-slab, full-SCR results. In \cref{fgr:TB-example-compare}(a), the full-SCR electrostatic energy curve is reproduced near the surface by $U_a(z)$ from the thin slab TB calculation. Deviations occur only near the leftmost atoms in the each thin slab (i.e., at the virtual face), at $-$60, $-$80 and $-$100 \A\, respectively, relative to the surface defect position. $z_d$ was found to be 1-2 \A\ left of the leftmost atoms in all cases. To the left of $z_d$, the remainder of the full-SCR curve is nicely reproduced by the numerical solution of \cref{eq:P-B,eq:P-Bpn} within the effective-mass approximation, at a much lower computational cost than an explicit treatment. CREST performs well at all three slab thicknesses examined, such that using a thicker slab simply amounts to accounting for more of the band-bending explicitly. This provides confirmation not only of the validity of an electrostatic treatment beyond a certain distance from the surface, but also of the assumption that the effect of the charged sheet on the electronic structure at the slab's virtual, passivated surface is negligible. Note that when less of the band-bending is accounted for explicitly in the slab, $\Delta Q$ increases and more charge is placed on the sheet. Despite this, a single value of $\Delta\varphi_{v}$, obtained from the fully-passivated reference calculation (see \cref{fgr:theory-FP}), is used successfully throughout. This validates our above-made assumption regarding the negligible effect of the SCR field on the electronic structure of the virtual surface.

Focusing now on the 30-site slab, the value of $U_d$ was found by CREST to be $\sim -$165 mV, relative to $U_l$ in the (right-side) vacuum. This 165 mV energy difference includes both the band bending explicitly included in the thin slab and the surface dipole (the sum of which is denoted by $e \Delta \varphi_S^{tot}$ in \cref{fgr:theory-main}(d)). The band bending in the effective-mass bulk was found by CREST to be $\Delta\varphi_b = -175~\text{mV}$, and the sum of this and $\Delta \varphi_S^{tot}$ thus reaches the full-SCR value of $-$340 mV. The DOS of the CREST slab matches the full-SCR curve at all energies from the surface state to about 0.15 eV below the VBM. The discrepancy at lower energies originates from bulk states that reside in the region far from the surface. These are included explicitly in the full-SCR calculation but not in the thin slab
\bibnote{
Our original full-SCR system contains 300 sites. In the absence of a SCR, the solution of the system Hamiltonian would result in 600 states that are delocalized throughout the slab. However, in the presence of a SCR, the bending of the bands upwards (to higher energies) towards the surface causes some states to localize nearer the surface and others to localize far from it. When simulating the surface using only 30 atomistic sites (i.e., 60 states), the states sacrificed in favor of an electrostatic description are those furthest from the surface. Accordingly, these states are missing from the DOS.
}.

In the CREST simulations in \cref{fgr:TB-example-compare}, the charged sheet was placed 12 \A\ left of the leftmost atoms. We emphasize that this is an arbitrary choice, which has no effect on the results shown (as long as the slab remains more than a few \A\ away from the slab's virtual face). This is so because owing to Gauss' law the field of an infinite, uniform, two-dimensional sheet of charge is constant. This observation has been confirmed numerically in our calculations (not shown for brevity).

\subsection*{Fully, partially and unpinned scenarios}
At a doped semiconductor surface, the equilibrium value of $E_F$, relative to the surface band gap, is affected by both the bulk and the surface. At one extreme, it may be entirely determined by the surface state, such that its value is independent of the bulk doping. We refer to this scenario as ``full Fermi-level pinning''. At the other extreme, the influence of the surface state may be negligible, and so $E_F$ matches its bulk value and no SCR forms. We refer to this scenario as an ``unpinned Fermi level''. The system of \cref{fgr:TB-example-fullSCR} displays an intermediate case: On the one hand, significant band bending is evident, such that $E_F$ relative to the surface band does not match its bulk value and the influence of the surface state is not negligible. On the other hand, the fact that the final $E_F$ resides \textit{above} the acceptor state level (see \cref{fgr:TB-example-fullSCR}(a) and \cref{fgr:TB-example-compare}(b)) indicates that it is significantly influenced by the bulk, because if it were not, $E_F$ would reside at or below the (empty) acceptor state energy. Thus, the Fermi level is neither unpinned nor fully-pinned. We refer to this scenario as ``partial'' or ``intermediate pinning''. We also use the terms ``strong pinning'' or ``weak pinning'' to indicate a deciding influence of the surface or bulk, respectively, on the position of $E_F$ at the surface.

\begin{figure}[htb!]
\centering
  \begin{tabular}{ l l }
    (a) & (b)\\
    \pstool![width=0.45\textwidth]{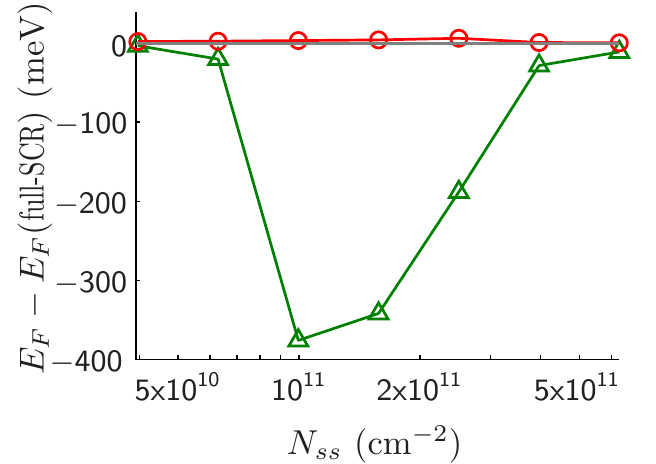}{
      \psfrag{x1}[t][t]{\setlength{\tabcolsep}{0pt}\begin{tabular}{c}
        $N_{ss}$ (cm$^{-2}$)
      \end{tabular}}
      \psfrag{x1t1}[t][t]{}
      \psfrag{x1t2}[t][t]{\small{$\mathsf{5x10^{10}}$}}
      \psfrag{x1t3}[t][t]{}
      \psfrag{x1t4}[t][t]{}
      \psfrag{x1t5}[t][t]{}
      \psfrag{x1t6}[t][t]{}
      \psfrag{x1t7}[t][t]{\small{$\mathsf{10^{11}}$}}
      \psfrag{x1t8}[t][t]{\small{$\mathsf{2x10^{11}}$}}
      \psfrag{x1t9}[t][t]{}
      \psfrag{x1t10}[t][t]{}
      \psfrag{x1t11}[t][t]{\small{$\mathsf{5x10^{11}}$}}
      \psfrag{x1t12}[t][t]{}
      \psfrag{y1}[b][b]{\setlength{\tabcolsep}{0pt}\begin{tabular}{c}
        $E_F - E_F$(\scalebox{0.75}[1.0]{full-SCR}) (meV)
      \end{tabular}}
      \psfrag{y1t1}[r][r]{\small{$\mathsf{-400}$}}
      \psfrag{y1t2}[r][r]{\small{$\mathsf{-300}$}}
      \psfrag{y1t3}[r][r]{\small{$\mathsf{-200}$}}
      \psfrag{y1t4}[r][r]{\small{$\mathsf{-100}$}}
      \psfrag{y1t5}[r][r]{\small{$\mathsf{0}$}}
    }&
    \pstool![width=0.45\textwidth]{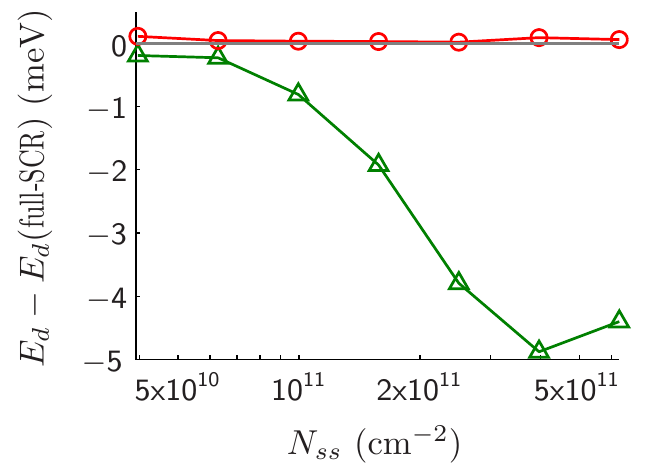}{
      \psfrag{x1}[t][t]{\setlength{\tabcolsep}{0pt}\begin{tabular}{c}
        $N_{ss}$ (cm$^{-2}$)
      \end{tabular}}
      \psfrag{x1t1}[t][t]{}
      \psfrag{x1t2}[t][t]{\small{$\mathsf{5x10^{10}}$}}
      \psfrag{x1t3}[t][t]{}
      \psfrag{x1t4}[t][t]{}
      \psfrag{x1t5}[t][t]{}
      \psfrag{x1t6}[t][t]{}
      \psfrag{x1t7}[t][t]{\small{$\mathsf{10^{11}}$}}
      \psfrag{x1t8}[t][t]{\small{$\mathsf{2x10^{11}}$}}
      \psfrag{x1t9}[t][t]{}
      \psfrag{x1t10}[t][t]{}
      \psfrag{x1t11}[t][t]{\small{$\mathsf{5x10^{11}}$}}
      \psfrag{x1t12}[t][t]{}
      \psfrag{y1}[b][b]{\setlength{\tabcolsep}{0pt}\begin{tabular}{c}
        $E_d - E_d$(\scalebox{0.75}[1.0]{full-SCR}) (meV)
      \end{tabular}}
      \psfrag{y1t1}[r][r]{\small{$\mathsf{-5}$}}
      \psfrag{y1t2}[r][r]{\small{$\mathsf{-4}$}}
      \psfrag{y1t3}[r][r]{\small{$\mathsf{-3}$}}
      \psfrag{y1t4}[r][r]{\small{$\mathsf{-2}$}}
      \psfrag{y1t5}[r][r]{\small{$\mathsf{-1}$}}
      \psfrag{y1t6}[r][r]{\small{$\mathsf{0}$}}
    }\\
    ~ & (c)\\
    ~ &
    \pstool![width=0.45\textwidth]{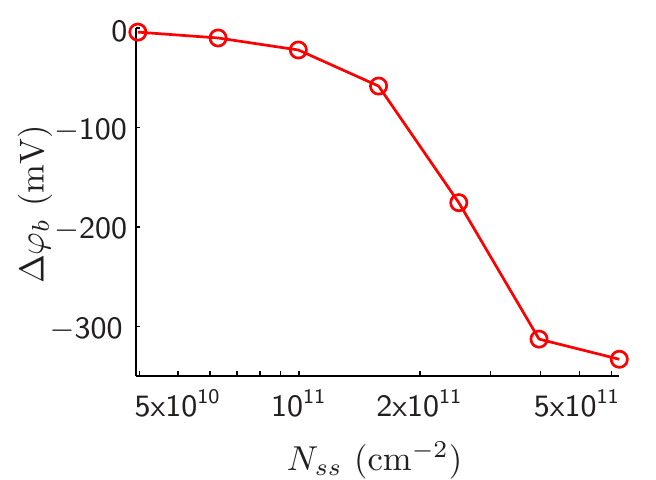}{
      \psfrag{x1}[t][t]{\setlength{\tabcolsep}{0pt}\begin{tabular}{c}
        $N_{ss}$ (cm$^{-2}$)
      \end{tabular}}
      \psfrag{x1t1}[t][t]{}
      \psfrag{x1t2}[t][t]{\small{$\mathsf{5x10^{10}}$}}
      \psfrag{x1t3}[t][t]{}
      \psfrag{x1t4}[t][t]{}
      \psfrag{x1t5}[t][t]{}
      \psfrag{x1t6}[t][t]{}
      \psfrag{x1t7}[t][t]{\small{$\mathsf{10^{11}}$}}
      \psfrag{x1t8}[t][t]{\small{$\mathsf{2x10^{11}}$}}
      \psfrag{x1t9}[t][t]{}
      \psfrag{x1t10}[t][t]{}
      \psfrag{x1t11}[t][t]{\small{$\mathsf{5x10^{11}}$}}
      \psfrag{x1t12}[t][t]{}
      \psfrag{y1}[b][b]{\setlength{\tabcolsep}{0pt}\begin{tabular}{c}
        $\Delta\varphi_b$ (mV)
      \end{tabular}}
      \psfrag{y1t1}[r][r]{\small{$\mathsf{-300}$}}
      \psfrag{y1t2}[r][r]{\small{$\mathsf{-200}$}}
      \psfrag{y1t3}[r][r]{\small{$\mathsf{-100}$}}
      \psfrag{y1t4}[r][r]{\small{$\mathsf{0}$}}
    }\\
  \end{tabular}
  \caption{Results from confined-SCR (green lines, triangles) and CREST (red lines, circles) calculations using our 1-D TB code, relative to the reference full-SCR calculation: (a) Fermi levels and (b) energy of the defect state, $E_d$. The horizontal line denotes 0, the reference value. Energies are referenced to the maximum of the bulk valence band, $E_{VBM}$, in each respective calculation. Panel (c) shows the values for $\Delta\varphi_b$ resulting from the CREST calculations. Bulk doping: $n$-type, $N_{dop} = 1.25 \times 10^{17} \text{cm}^{-3}$.}
  \label{fgr:TB-all-vs-full}
\end{figure}

We are able to explore different pinning regimes within the TB model by varying $N_{ss}$ in the system while keeping $N_{dop}$ constant. In \cref{fgr:TB-all-vs-full}, SCR-confined and CREST results throughout the range from no- to full-pinning are referenced to the full-SCR calculation, focusing on the values obtained for $E_F$ and for the defect level in the gap, $E_d$, relative to the bulk valence band maximum in each respective calculation. The values of $\Delta\varphi_b$ obtained from the CREST calculations are also shown (\cref{fgr:TB-all-vs-full}(c)), indicating the remaining band bending which is not included explicitly in the atomistic slab. Results for a $p$-type bulk with similar $N_{dop}$ and a deep donor level are very similar, but all errors are reversed in sign, as are $\Delta\varphi_b$ values in \cref{fgr:TB-all-vs-full}(c) (not shown for brevity).

The progression from no to full $E_F$ pinning, corresponding to about a tenfold increase in $N_{ss}$, is clear in the magnitude of the errors in $E_F$ due to SCR confinement (\cref{fgr:TB-all-vs-full}(a)). In the extreme case of no pinning, the ``doped-atom'' approach is sufficient, as also seen in Ref. \cite{Sinai2013} using DFT. In the other extreme, the error in $E_F$ of the confined-SCR system is also small, as $E_F$ becomes pinned to the defect state and becomes insensitive to the bulk. In the middle of the range, however, charge drawn from the bulk-reservoir becomes a deciding factor in the final $E_F$. This leads to errors of as much as 0.4 eV in $E_F$ of the confined-SCR slab, almost half of the semiconductor gap of $\sim$1 eV.

The errors in the energy of the surface state (\cref{fgr:TB-all-vs-full}(b)) follow a different progression, since they are determined by the SCR-derived field. This field becomes greater the more charge is transferred to the surface, which in turn is associated with stronger pinning. In the SCR-confined case this charge transfer is curtailed by the limited system size.

As \cref{fgr:TB-all-vs-full} shows, in all these cases CREST corrects nicely for the errors introduced by the thin slab setup. The $\Delta\varphi_b$ values obtained, shown in \cref{fgr:TB-all-vs-full}(c), follow the magnitude of charge transfer $\Delta Q$ from the bulk, which becomes greater as $E_F$ is more strongly pinned. This is precisely the charge ``missing'' from the confined-SCR slab.

\begin{figure}[H]
\centering
  \begin{tabular}{ l }
    \pstool![width=0.7\textwidth]{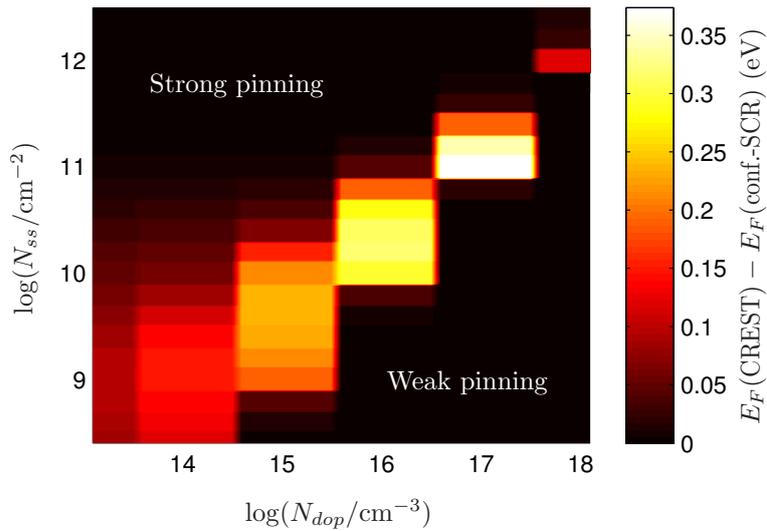}{
      \psfrag{x1}[t][t]{\setlength{\tabcolsep}{0pt}\begin{tabular}{c}
        $\log(N_{dop}/\text{cm}^{-3})$
      \end{tabular}}
      \psfrag{y1}[b][b]{\setlength{\tabcolsep}{0pt}\begin{tabular}{c}
        $\log(N_{ss}/\text{cm}^{-2})$
      \end{tabular}}
      \psfrag{y2}[c][c]{\setlength{\tabcolsep}{0pt}\begin{tabular}{c}
        $E_F($\scalebox{0.9}[1.0]{CREST}$) - E_F($\scalebox{0.9}[1.0]{conf.-SCR}$)$ (eV)
      \end{tabular}}
      \psfrag{lb1}[c][c]{\setlength{\tabcolsep}{0pt}\begin{tabular}{c}
        \textcolor[rgb]{1,1,1}{Strong pinning}
      \end{tabular}}
      \psfrag{lb2}[c][c]{\setlength{\tabcolsep}{0pt}\begin{tabular}{c}
        \textcolor[rgb]{1,1,1}{Weak pinning}
      \end{tabular}}
    }\\
  \end{tabular}
  \caption{CREST corrections to $E_F$ values in thin-slab TB simulations of SCR-possessing systems. Energies are referenced to maximum of the bulk valence band, $E_{VBM}$, in each respective calculation. All doping densities shown are $n$-type, and the surface state is a deep acceptor, approx. 0.1 eV above the $E_{VBM}$.}
  \label{fgr:TB-sys-scan}
\end{figure}

We conclude this section with an examination of CREST corrections to a broad array of thin-slab systems with various doping rates and surface state densities, focusing on corrections to $E_F$. This is shown in \cref{fgr:TB-sys-scan} for $n$-type doping. Very similar results are obtained from $p$-type systems, albeit with a reversed sign of the corrections (not shown for brevity). These results demonstrate the interplay between bulk and surface that determines the position of $E_F$ at the surface. The different pinning regimes are clearly recognizable in \cref{fgr:TB-sys-scan}. The top-left area, corresponding to high values of $N_{ss}$ and low values of $N_{dop}$, is the strong-pinning regime, and the bottom-right is the weak-pinning regime. In both of these the corrections of CREST to the $E_F$ of the confined-SCR slab are negligible, as also seen at the left and right of \cref{fgr:TB-all-vs-full}(a), which corresponds to ``travelling'' vertically in \cref{fgr:TB-sys-scan}. However, in the intermediate-pinning regime, appearing along the bottom-left to top-right diagonal, CREST provides large corrections amounting to significant fractions of the forbidden gap energy.

\section{Density functional theory calculations}
\label{sec:DFT}
In order to use CREST with first-principles DFT calculations, we modified the FHI-AIMS code \cite{Blum2009} to allow for the inclusion of point-charges. A grid of these charges, periodic in parallel to the surface and at some distance from the virtual surface of the slab, can perfectly simulate the electrostatic effect of a fixed charged sheet, because the potential of a periodic point-charge arrangement in the normal direction has a constant leading term, i.e., of a uniform sheet, and all other terms decay exponentially with a maximal decay length of $a / 2\pi$, where $a$ is the point-charge grid spacing \cite{Lennard-Jones1928}. Thus at a distance reasonably larger than $a / 2\pi$, no additional fields due to the discretization of the sheet will remain, leaving only the field of a uniform sheet. Because FHI-AIMS is an all-electron code, ``doped'' atoms are naturally included by a simple modification of the nuclear and valence charges of the bulk atoms, as has been done in previous work \cite{Richter2013, Xu2013, Moll2013, Hofmann2013}. We use throughout the generalized-gradient approximation to the exchange-correlation functional, given by Perdew, Burke and Ernzerhof (PBE functional) \cite{Perdew1996}. In FHI-aims, the basis set is hierarchically ordered into tiers. Here, geometries were relaxed with the ``light'' defaults for integration grids and basis functions as provided by FHI-aims, which correspond to a tier-1 basis set. Electronic structures were subsequently determined by single-point calculations with ``tight'' defaults, which are based on a tier-2 basis set. Calculations were always performed with periodic boundary conditions in all three dimensions, with a dipolar correction \cite{Neugebauer1992, Makov1995, Bengtsson1999} applied in the vacuum between the real surface and the charged sheet in the next periodic image, to prevent the appearance of a spurious field in this region. It is important not to confuse this spurious field with the intentional field in the vacuum between the charged sheet and the virtual surface. We used Monkhorst-pack \cite{Monkhorst1976c} k-point grids centered on the $\Gamma$-point, with 20x20x1 and 10x20x1 points for (1x1) and (2x1) surfaces, respectively. Eigenstate occupations were smeared with a Fermi-Dirac distribution at room temperature (RT, $298.15~K$; note, however, that this is not a requirement for the use of CREST in general). 

In the results shown below, we considered simulation slabs containing 26 Si layers ($\sim$ 40 \A\ thick), constructed as follows: The relaxed Si bulk (with the lattice constant found to be 5.47 \A) was cleaved in the (111) direction and capped on both sides with hydrogen atoms. The outer 6 Si layers and the H atoms on each side were then allowed to relax, leading to a fully-passivated slab. In calculations of subsequent unpassivated surfaces, only the topmost 10 Si layers were replaced and allowed to relax, with the center and the bottom (virtual) side frozen. The 10 topmost Si layers' geometry was preserved in the 20-layer slabs, while in the 14-layer slabs, only the upper 6 Si layers were relaxed. The starting geometry of the clean-cleaved Si(111) surface (introduced below) was obtained by cleavage of the bulk. The starting geometry of the positively-buckled Si(111) (2x1) reconstruction was based on LEED data obtained by Himpsel et al \cite{Himpsel1984}, and of the negatively-buckled Si(111) (2x1) reconstruction on recently published computational data \cite{Violante2014} (these surfaces are also introduced below). All relaxations were performed before the application of CREST. For the effective-mass treatment, well-established values of the Si bulk parameters were taken from the literature (Ref. \cite{Sze2006}), namely $\varepsilon = 11.9$, $N_V = 2.65 \times 10^{19} \text{cm}^{-3}$ and $N_C = 2.80 \times 10^{19} \text{cm}^{-3}$. These adequately describe bulk Si calculated with PBE, as demonstrated in a previous comparison of ``doped-atom'' PBE to effective-mass results using these values \cite{Sinai2013}. Values of $E_g$ = 0.70 - 0.71 eV were extracted from fully-passivated system calculations as described above (see \cref{fgr:theory-FP}), as were values of $\Delta\varphi_{v}$ = (-1.0)-(-1.1) eV and of the work function on the virtual side, $\Phi_v = \Delta\varphi_{v} + W_b$ = 4.3, 4.1 and 4.0 eV for intrinsic, $10^{17}$ and $2 \times 10^{19} \text{cm}^{-3}$ $n$-type doping, respectively. The convergence parameter for CREST was set to 1 meV on the Fermi level and band bending. The CREST self-consistent cycle converged within at most 7 iterations in all cases.

We demonstrate CREST/DFT by applying the technique to three different reconstructions of the Si(111) surface, focusing on the electronic structure. The first of these is the ``clean-cleaved'' (1x1) surface (denoted henceforth by ``CC''). Cleavage along the (111) plane of Si usually occurs through the middle of the covalent bond pointing along the (111) direction \cite{Srivastava1997}. This leaves dangling bonds, where the surface atoms are under-coordinated relative to the bulk. These dangling bonds are half-filled surface states, causing Fermi-level pinning and charge transfer to the surface, the magnitude of which depends on the bulk doping. The second and third surfaces examined are based on the (2x1) reconstruction, which appears when the cleaving process of the Si(111) surface is performed at high vacuum and room temperature. The commonly accepted model for this reconstruction involves a $\pi$-bonded chain structure suggested by Pandey \cite{Pandey1981, Northrup1982, Srivastava1997}. An improved model indicated that the chains can buckle \cite{Himpsel1984}. This buckling may occur in two directions, leading to the existence of two isomers, termed ``positive'' and ``negative'' \cite{Rohlfing2000, Bussetti2011}, and denoted below by ``PB'' and ``NB'', respectively. For each of these surfaces, we consider three representative doping concentrations, namely intrinsic, $n$-type with $N_{dop} = 10^{17} \text{cm}^{-3}$ (moderately doped, MDn) and $n$-type with $N_{dop} = 2 \times 10^{19} \text{cm}^{-3}$ (highly doped, HDn).

\begin{figure}[H]
\centering
  \begin{tabular}{ l }
    \pstool![width=0.7\textwidth]{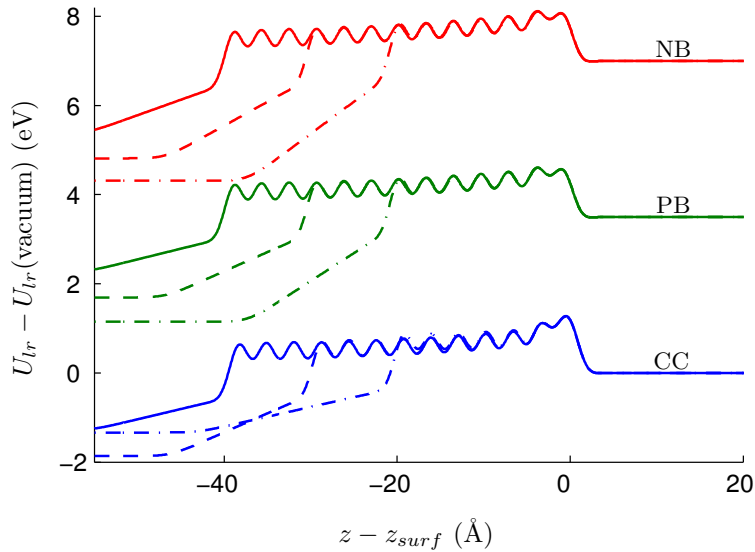}{
      \psfrag{x1}[t][t]{\setlength{\tabcolsep}{0pt}\begin{tabular}{c}
        $z - z_{surf}$ (\A)
      \end{tabular}}
      \psfrag{y1}[b][b]{\setlength{\tabcolsep}{0pt}\begin{tabular}{c}
        $U_{lr} - U_{lr}(\text{vacuum})$ (eV)
      \end{tabular}}
      \psfrag{lb1}[c][c]{\setlength{\tabcolsep}{0pt}\begin{tabular}{c}
        \small{CC}
      \end{tabular}}
      \psfrag{lb2}[c][c]{\setlength{\tabcolsep}{0pt}\begin{tabular}{c}
        \small{PB}
      \end{tabular}}
      \psfrag{lb3}[c][c]{\setlength{\tabcolsep}{0pt}\begin{tabular}{c}
        \small{NB}
      \end{tabular}}
    }\\
  \end{tabular}
  \caption{Development of the plane-averaged long-range electrostatic energy, $U_{lr}$, as obtained from FHI-AIMS, in CREST simulations with slabs comprising 26 (solid lines), 20 (dashed lines) and 14 (dash-dotted lines) Si layers, of the highly-doped (HDn) CC (blue lines), PB (green lines) and NB (red lines) surfaces. Curves are aligned horizontally to $z_{surf}$, the position of the topmost surface atom in each calculation, and vertically to $U_{lr}$ in the vacuum outside (to the right of) the real surface examined. For clarity, the curves for the PB surface are offset upward by 3.5 eV, and for the NB surface by 7 eV.}
  \label{fgr:DFT-conv}
\end{figure}

As a preliminary step, we performed CREST/DFT calculations of the examined surfaces and doping levels with slabs comprising 26, 20 and 14 Si layers. The plane-averaged long-range electrostatic energy (denoted by $U_{lr}$) curves for the three surfaces examined in the HDn case are shown in \cref{fgr:DFT-conv}. As in the results shown in \cref{fgr:TB-example-compare}(a) above, it is clear that in general, the use of a thicker slab (identifiable by potential energy oscillations) amounts to inclusion of more of the band bending explicitly within the slab rather than in the effective-mass bulk. Accordingly, the thicker the slab, the smaller the charge on the sheet, as is evident from the slope of $U_{lr}$ in the vacuum to the left of the slab in each case. This trend is broken only in the thinnest CC surface calculation, which also displays artifacts within the slab, indicating that it is not sufficiently converged at 14 Si layers. The PB and NB calculations converge more quickly, and regardless, all surfaces are well converged at our working thickness of 26 Si layers. The surface work functions obtained from the two thicker slab calculations were equal within 0.01 eV in all systems (in most cases within 1-2 meV), and even results from the 14-layer slabs did not deviate from the thickest slabs by more than $\sim$0.02 eV. Total energy differences between the slabs were similarly converged, with $\sim$0.01 meV at most between the thickest slabs. As an additional test, we carried out a calculation of the CC surface of HDn Si, using ``doped'' atoms alone, with a deep (98 Si layers, $\sim$15 nm thick) slab. This thickness is sufficient to capture almost the entire SCR in this system due to its relatively small width, though the computation is demanding. It was confirmed that the results of the corresponding CREST calculation indeed match this ``full-SCR'' calculation (not shown for brevity).

Taken together, these results, analogously to the TB system results in \cref{fgr:TB-example-compare}, affirm the applicability of an electrostatic treatment within a short distance from the surface; support the robustness of the virtual surface's electronic structure and potential drop against realistic changes in the sheet charge (note that the largest fields occur in the HDn cases shown); and demonstrate that the CREST results are converged at slab thicknesses comparable to those required for the convergence of other surface properties. We therefore proceed to examine the effects of doping on the band structures of the CC and PB surfaces. Results for the NB surface are omitted henceforth for brevity, as the trends appearing there are intermediate between those appearing in the band structures discussed below.

\begin{figure}[htb!]
\centering
  \begin{tabular}{ m{0.03\textwidth} m{0.45\textwidth} m{0.45\textwidth} }
    ~ &
    \multicolumn{1}{c}{\large{\textbf{With CREST}}}&
    \multicolumn{1}{c}{\large{\textbf{Without CREST}}}\\
    ~ & (a) & (b)\\
    \rotatebox{90}{\large{\textbf{CC (1x1)}}} &
    \pstool![width=0.45\textwidth]{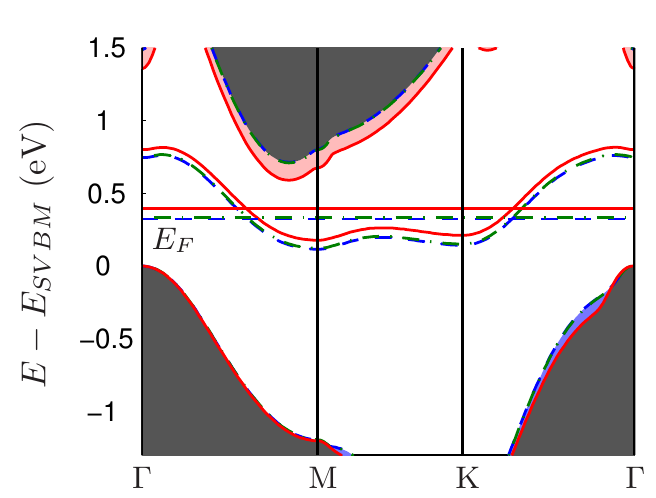}{
      \psfrag{y1}[b][b]{\setlength{\tabcolsep}{0pt}\begin{tabular}{c}
        $E - E_{SVBM}$ (eV)
      \end{tabular}}
      \psfrag{G}{\small{$\Gamma$}}
      \psfrag{M}{\small{M}}
      \psfrag{K}{\small{K}}
      \psfrag{lb1}[l][l]{\small{$E_F$}}
    }&
    \pstool![width=0.45\textwidth]{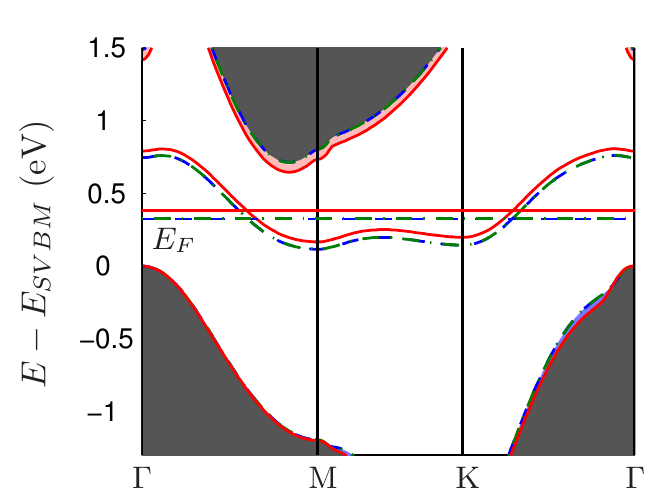}{
      \psfrag{y1}[b][b]{\setlength{\tabcolsep}{0pt}\begin{tabular}{c}
        $E - E_{SVBM}$ (eV)
      \end{tabular}}
      \psfrag{G}{\small{$\Gamma$}}
      \psfrag{M}{\small{M}}
      \psfrag{K}{\small{K}}
      \psfrag{lb1}[l][l]{\small{$E_F$}}
    }\\
    ~ & (c) & (d)\\
    \rotatebox{90}{\large{\textbf{PB (2x1)}}} &
    \pstool![width=0.45\textwidth]{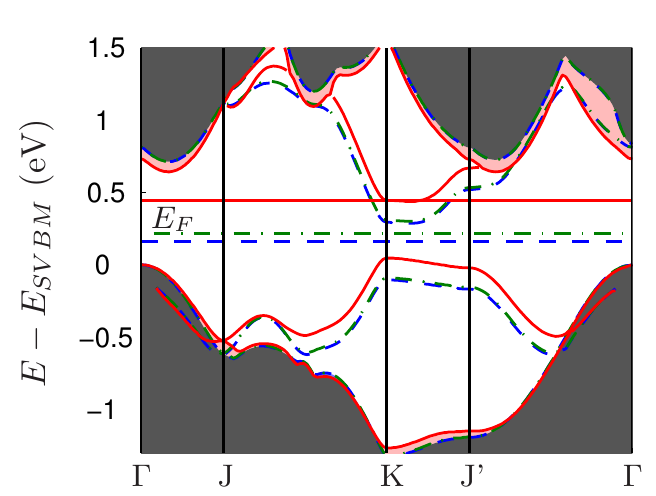}{
      \psfrag{y1}[b][b]{\setlength{\tabcolsep}{0pt}\begin{tabular}{c}
        $E - E_{SVBM}$ (eV)
      \end{tabular}}
      \psfrag{G}{\small{$\Gamma$}}
      \psfrag{J}{\small{J}}
      \psfrag{J'}{\small{J'}}
      \psfrag{K}{\small{K}}
      \psfrag{lb1}[bl][bl]{\small{$E_F$}}
    }&
    \pstool![width=0.45\textwidth]{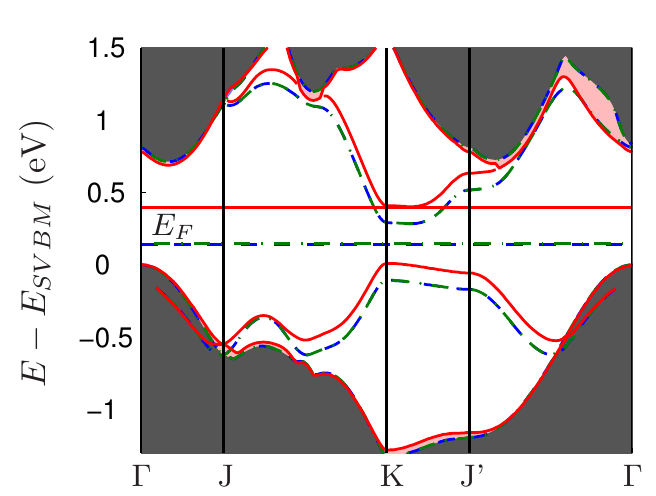}{
      \psfrag{y1}[b][b]{\setlength{\tabcolsep}{0pt}\begin{tabular}{c}
        $E - E_{SVBM}$ (eV)
      \end{tabular}}
      \psfrag{G}{\small{$\Gamma$}}
      \psfrag{J}{\small{J}}
      \psfrag{J'}{\small{J'}}
      \psfrag{K}{\small{K}}
      \psfrag{lb1}[bl][bl]{\small{$E_F$}}
    }\\
    ~ & (e) & ~\\
    ~ &
\begingroup%
  \setlength{\unitlength}{0.4\textwidth}
  \begin{picture}(1,0.74896464)%
    \put(0,0){\includegraphics[width=\unitlength]{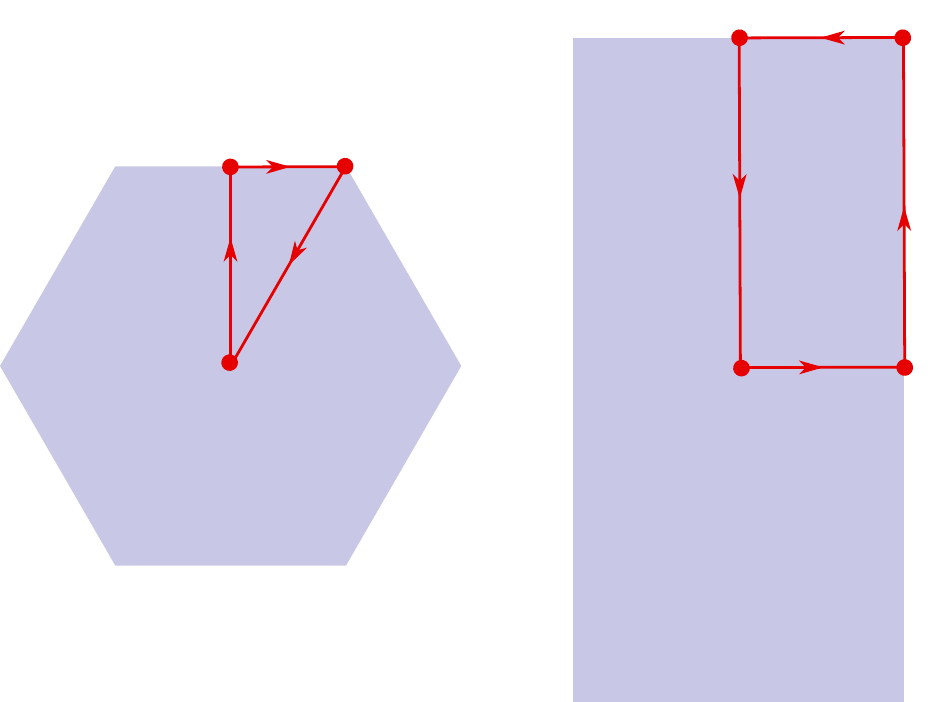}}%
    \put(0.22,0.59){\color[rgb]{0,0,0}\makebox(0,0)[lb]{\smash{M}}}%
    \put(0.38,0.59){\color[rgb]{0,0,0}\makebox(0,0)[lb]{\smash{K}}}%
    \put(0.23,0.30){\color[rgb]{0,0,0}\makebox(0,0)[lb]{\smash{$\Gamma$}}}%
    \put(0.74,0.30){\color[rgb]{0,0,0}\makebox(0,0)[lb]{\smash{$\Gamma$}}}%
    \put(0.98,0.30){\color[rgb]{0,0,0}\makebox(0,0)[lb]{\smash{J}}}%
    \put(0.74,0.72){\color[rgb]{0,0,0}\makebox(0,0)[lb]{\smash{J'}}}%
    \put(0.98,0.72){\color[rgb]{0,0,0}\makebox(0,0)[lb]{\smash{K}}}%
  \end{picture}%
\endgroup%
    &
    \textbf{Legend:}\linebreak
    \linebreak
\begingroup%
  \setlength{\unitlength}{0.4\textwidth}
  \begin{picture}(1,0.57321958)%
    \put(0,0){\includegraphics[width=\unitlength]{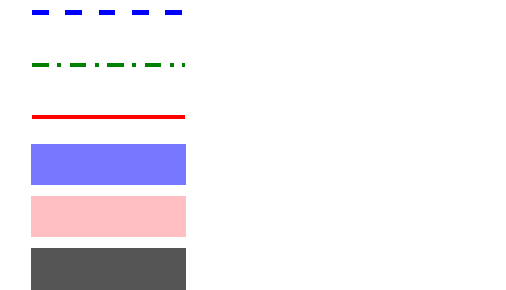}}%
    \put(0.43074057,0.22281799){\color[rgb]{0,0,0}\makebox(0,0)[lb]{\smash{Only intrinsic / MDn}}}%
    \put(0.43074057,0.1197048){\color[rgb]{0,0,0}\makebox(0,0)[lb]{\smash{Only HDn}}}%
    \put(0.4340242,0.01701011){\color[rgb]{0,0,0}\makebox(0,0)[lb]{\smash{All doping densities}}}%
    \put(0.42777887,0.52602542){\color[rgb]{0,0,0}\makebox(0,0)[lb]{\smash{Intrinsic}}}%
    \put(0.42864806,0.31941229){\color[rgb]{0,0,0}\makebox(0,0)[lb]{\smash{HDn}}}%
    \put(0.42903433,0.42252576){\color[rgb]{0,0,0}\makebox(0,0)[lb]{\smash{MDn}}}%
  \end{picture}%
\endgroup%
    \\
  \end{tabular}
  \caption{Surface band structures of the metallic, clean-cleaved Si(111)(1x1) surface (a \& b), and of the semiconducting, positively-buckled $\pi$-chain Si(111)(2x1) surface (c \& d). Results from both CREST-corrected (a \& c) and confined-SCR (b \& d) calculations are shown. Within each panel, different doping levels are compared: Intrinsic (blue dashed lines), moderately $n$-doped (MDn; $N_{dop} = 10^{17} \text{cm}^{-3}$, green dot-dashed lines) and heavily $n$-doped (HDn; $N_{dop} = 2 \times 10^{19} \text{cm}^{-3}$, red solid lines). $E_F$ values are marked by straight horizontal lines. The surface bands are the free-standing (curved) lines. The ``bulk-like'' bands (see text) are shaded, with blue indicating regions where bands exist only in the intrinsic/MDn cases, light red where bands exist only in the HDn case, and dark gray where bands exist at all doping densities. The edges are marked with the respective lines for clarity. Energies are referenced to the maximum of the highest occupied bulk-like band, $E_{SVBM}$, in each respective surface band diagram. Panel (e) describes the paths taken in k-space.}
  \label{fgr:Si111-bs}
\end{figure}

The left-hand panels of \cref{fgr:Si111-bs} describe the surface band-structure diagrams, at all three representative doping rates, of the CC (a) and PB (c) surface reconstructions, respectively, as calculated with CREST. The corresponding confined-SCR calculations without CREST are shown for the CC and PB surfaces in panels (b) and (d), respectively. In these diagrams, the majority of states in the vicinity of $E_F$ form continuous energy bands which are ``bulk-like'', i.e., they resemble the bulk bands of Si, consistent with the rapid decay of the surface states away from the surface. The surface states are clearly identifiable as isolated states within the bulk-like gap. As expected, the CC surface is metallic, with $E_F$ positioned within the surface band at all doping rates, while the PB surface is semiconducting. The PB surface band structure agrees qualitatively with that obtained from many body-perturbation theory \cite{Northrup1991, Bussetti2011}, and even though the PBE approximation causes a significant underestimation of the gap, the salient features are captured well enough for the purposes of our demonstration. All energies in \cref{fgr:Si111-bs} (e.g., $E_F$) are given relative to the maximum of the highest occupied bulk-like band, denoted by $E_{SVBM}$, in each respective system. The amount of charge transferred to each surface as calculated by CREST, $\Delta Q$, and the ``missing'' band bending CREST accounts for, $\Delta\varphi_b$, are collected in \cref{tbl:Si111-bbsc}. This table also includes an estimation of the total band bending in the system (denoted by $\Delta\varphi_{BB}^{tot}$ in \cref{fgr:theory-main}(a)), which is the sum of $\Delta\varphi_b$ and the potential drop occurring inside the slab, from a point where the short-range surface oscillations in the macroscopic electrostatic energy have visibly decayed (around the 4th or 5th Si layer in from the surface).

\begin{table}[htb!]
  \caption{CREST results for Si(111) surfaces}
  \centering
  \begin{tabular}{l c c c}
    \hline
    ~ & \multicolumn{3}{c}{Clean-cleaved (1x1)}\\
    \hline
    $N_{dop} (\text{cm}^{-3})$ & $\Delta Q (\text{e/cm}^{-2})$ & $\Delta\varphi_b$ (V)& $\Delta\varphi_{BB}^{tot}$ (V, est.)\\
    \hline
    Intrinsic & $-1.61 \times 10^{9}$ & -0.013 & (-0.07)\\
    $n$-type $10^{17}$ & $-4.63 \times 10^{11}$ & -0.189 & (-0.26)\\
    $n$-type $2 \times 10^{19}$ & $-2.55 \times 10^{12}$ & -0.049 & (-0.36)\\
    \hline
    \multicolumn{4}{c}{~}\\
    ~ & \multicolumn{3}{c}{Positive buckling (2x1)}\\
    \hline
    $N_{dop} (\text{cm}^{-3})$ & $\Delta Q (\text{e/cm}^{-2})$ & $\Delta\varphi_b$ (V)& $\Delta\varphi_{BB}^{tot}$ (V, est.)\\
    \hline
    Intrinsic & $-1.21 \times 10^{11}$ & -0.186 & (-0.22)\\
    $n$-type $10^{17}$ & $-6.13 \times 10^{11}$ & -0.311 & (-0.36)\\
    $n$-type $2 \times 10^{19}$ & $-2.65 \times 10^{12}$ & -0.052 & (-0.35)\\
    \hline
  \end{tabular}
  \label{tbl:Si111-bbsc}
\end{table}

We start by discussing the band structures at the CC surface (\cref{fgr:Si111-bs}(a)). At all three doping densities, the metallic nature of the surface is evident in the fact that $E_F$ (horizontal line) passes through the surface band (curved line) regardless of doping, with negligible changes in its position relative to the band. $E_F$ is therefore strongly pinned at the surface. As a result of this, and as is also clear from \cref{fgr:Si111-bs}(a), the change of the bulk doping density from intrinsic (blue lines) to MDn (green) has only minute effects on the band structure diagram. We do, however, expect it to have significant effects on the bulk (which are accounted for by CREST), as indeed is evident from the increase in the band bending occurring in the semi-infinite bulk, from $\sim$-0.01 V in the intrinsic case to $\sim$-0.17 V in the MDn case (\cref{tbl:Si111-bbsc}, upper part). In keeping with this order-of magnitude increase in $|\Delta \varphi_b|$, the magnitude of the charge transferred to the surface, $|\Delta Q|$, is increased by two orders of magnitude. The total band bending also follows the same progression. Note that the exact numerical value of $\Delta\varphi_{BB}^{tot}$ depends on the somewhat arbitrary distinction between surface dipole and band-bending, i.e., on the extent of the region where the potential is considered as a ``short-range surface effect''. This is especially so for the intrinsic case, where this may cause most of the (overall very small) band bending to seem to occur in the slab.

In contrast to the MDn case, in the HDn case (red lines) actual changes in the band structure diagram occur relative to the intrinsic or MDn cases. First, the surface band is shifted up by $\sim$0.07 eV, due to the field originating from the SCR. This effect is exactly the same as seen in the TB model system in \cref{fgr:TB-all-vs-full}(b) at high $N_{ss}$ values. Note that because $E_F$ remains pinned to the surface band, it too increases relative to the bulk-like bands in the slab (monochrome shaded regions in \cref{fgr:Si111-bs}(a)), and of course relative to the bulk itself (which is treated only with the effective mass approximation). The other significant change is that much of the band bending, which previously occurred outside the slab (and was accounted for by CREST), now occurs within it. Because the bands are bent up towards the surface, the highest occupied bulk-like band in the slab originates close to the surface, while the lowest unoccupied bulk-like band originates far from it. Thus, the inclusion of more band bending explicitly in the slab lowers the unoccupied bulk-like bands by $\sim$0.1 eV in \cref{fgr:Si111-bs}(a). This is accompanied by a decrease of $|\Delta \varphi_b|$ by approximately the same amount, to a few tens of mV (\cref{tbl:Si111-bbsc}, upper part). Note that despite this, CREST still accounts for a significant amount of ``missing'' charge transferred to the slab, an order of magnitude greater than in the MDn case.

We compare CREST results for the CC surface, \cref{fgr:Si111-bs}(a), with the band structure diagram in \cref{fgr:Si111-bs}(b), which is obtained for the same surface but in a ``confined-SCR'' configuration, i.e., without the use of CREST. A side-by-side comparison of the diagrams shows that they are quite similar, in particular at intrinsic or moderate doping. This is expected, in light of the strong pinning of $E_F$ which characterizes this surface. As also demonstrated in the TB model system (see \cref{fgr:TB-sys-scan}), in strongly-pinned cases even the use of ``doped'' atoms alone in a confined-SCR configuration captures correctly the position of $E_F$ relative to the bands at or very near the surface. Despite this, some differences do appear at high doping, the most noticeable of which is in the position of the unoccupied bulk-like band, which in the CREST-corrected system (\cref{fgr:Si111-bs}(a)) is visibly lower than in the confined-SCR system (\cref{fgr:Si111-bs}(b)). This is a direct result of the confinement, as the bands are forced to ``level out'' too early, similarly to the situation shown in the TB example in \cref{fgr:TB-example-compare}(a).

The results for the CC surface show that the use of CREST provides us with some insights as to the influence of the bulk doping even in a fully-pinned scenario. In particular, we obtain information about surface charging and the bulk potential drops. Importantly, note that while we were able to guess, due to pinning, that the slab band structure would be fairly well described with doped atoms (confined-SCR arrangement) alone, by the use of CREST we obtain this \textit{as a result of the calculation}, with \textit{no a priori} assumptions.

Turning to the CREST calculations of the PB surface (\cref{fgr:Si111-bs}(c)), we focus first on the intrinsic band structure (blue lines and blue/gray shaded regions). It is obvious that the system is semiconducting: $E_F$ resides within a forbidden gap between the maximum of the highest occupied band (HOB) and the minimum of the lowest unoccupied band (LUB). However, the gap is heterogeneous: The HOB maximum belongs to a bulk-like band (shown by the lower gray/blue shaded regions), while the LUB minimum belongs to a surface state (blue solid line). The lowest unoccupied bulk-like band minimum, found along the high-symmetry line between the $\Gamma$ and J points, is much higher. As a result, the $E_F$ position is determined by a number of factors including the bulk and surface DOS near the valence and conduction band edges respectively, the bulk doping, and the band bending in the bulk. This interplay points to a situation of partial pinning. Another result of this situation is clear from the significant amount of charge transfer to the slab and the bulk band bending accounted for in CREST (\cref{tbl:Si111-bbsc}, lower part): Already at intrinsic doping, a significant amount of charge resides on the surface, accompanied by a band bending of $\Delta \varphi_b = -0.18 \text{ V}$ occurring in the effective-mass bulk (of the $\sim -0.22 \text{ V}$ total).

Because $E_F$ in the PB surface is partially pinned, we see a noticeable up-shift of $\sim$0.06 V in $E_F$ in the MDn case relative to the intrinsic case, such that it comes to reside $\sim$0.08 V from the LUB minimum - approximately at the rule-of-thumb limit for degenerate doping, $3kT$. The surface bands, however, are shifted only slightly due the increase in the SCR field, while the bulk-like bands do not shift at all. The $|\Delta \varphi_b|$ accounted for by CREST is almost doubled (with a corresponding increase in $|\Delta\varphi_{BB}^{tot}|$), and $|\Delta Q|$ is 5-6 times larger. If we now consider the HDn case, we see some similar trends to those we saw in the HDn CC surface. The surface bands are higher by 0.15 V, $E_F$ is pinned to the LUB minimum, and the inclusion of much of the band bending inside the slab causes the unoccupied bulk-like band to lower by $\sim$0.1 eV, while $|\Delta \varphi_b|$ decreases to $\sim$0.05 eV. Other small changes also occur in the bulk-like bands. One interesting result in this context is the fact that due to the up-shifting of the occupied surface band, its energy maximum becomes higher than that of the highest occupied bulk-like band by about 0.05 eV, changing the character of the system's forbidden gap to being a pure surface gap. Also of interest is the fact that the total band bending does not increase significantly. This is a result of the partial pinning situation: $E_F$ is driven up relative to the surface bands, and together with the up-shift of the surface bands themselves, this preserves the energy difference between the bulk- and the surface-determined $E_F$ values despite the increase in the former.

Finally, we compare the PB surface CREST result, \cref{fgr:Si111-bs}(c), with an uncorrected, confined-SCR result for the same surface, \cref{fgr:Si111-bs}(d). In the confined-SCR system, upon going from intrinsic to moderate doping, as in the strongly-pinned CC surface (\cref{fgr:Si111-bs}(b)), hardly any change to the band structure is observed. However, while in the CC case, our use of CREST was able to validate this result, in the partially-pinned PB surface this result turns out to be erroneous, as charge transferred from the bulk in fact causes $E_F$ to shift up noticeably (\cref{fgr:Si111-bs}(c)). This is again in keeping with the trends observed in the TB model system for surfaces exhibiting partial $E_F$-pinning (see \cref{fgr:TB-sys-scan}). As for the HDn PB surface, we note that all of the phenomena highlighted in \cref{fgr:Si111-bs}(c), and in particular the bulk-like LUB down-shift, are underestimated in the confined-SCR HDn result. All of this confirms not only that the insights gained from our TB analysis are indeed valid when using DFT, but also that for an entirely correct description of electronic structure at charged semiconductor surfaces, a fully self-consistent treatment is needed. We further note that while here the use of CREST amounted to a quantitative correction, lack of proper treatment of SCR effects may also lead to qualitative failures, e.g., in the description of charge transfer, adsorption energies and surface work functions at an F4TCNQ-adsorbed ZnO surface \cite{Xu2013}.

\section{Summary}
We have presented an approach that allows the global effects of doping to be accounted for in simulations of the surfaces of semiconductors. The approach is based on the inclusion of a sheet of fixed charge within a periodic slab calculation, such that the overall net charge of the system is zero. The fixed charged sheet reproduces the electrostatic effects of the SCR, and the associated non-fixed charge duplicates the charge transferred from the doped semiconductor bulk to the surface (whereby the SCR is formed). The correct amount of charge is calculated based on the twin requirements of charge conservation and Fermi-level equilibration between the semi-infinite bulk, which is treated semi-classically, and the slab, which can be treated with atomistic detail. We call this technique CREST - the Charge-Reservoir-Electrostatic-Sheet Technique. CREST is implemented as a wrapper program around standard electronic structure codes.

We applied CREST first in conjunction with a simple tight-binding electronic-structure code. Using this, we demonstrated the ability of CREST to correct the errors introduced by the use of calculation cells that truncate the SCR, thus reproducing the results from much more computationally demanding calculations wherein the full SCR was included. This was demonstrated in a range of cases, from full- to partial- to no Fermi level pinning.

Finally, we used CREST with the DFT electronic structure code FHI-AIMS to treat the clean-cleaved, metallic (1x1) Si(111) surface, as well as the positively-buckled $\pi$-chain (or Pandey) (2x1) semiconducting rearrangement of the Si(111) surface. We calculated the magnitudes of the charging and band bending at these surfaces, and presented unique information about the doping-dependence of the surface band structures and the Fermi levels at these surfaces. This information demonstrates that these surfaces represent different degrees of $E_F$-pinning.

We expect CREST to be useful in wide range of semiconductor surface and interface simulations, and hope that the design of CREST as a freestanding procedure external to the electronic structure code will facilitate its ready application. Beyond examples we have described in the introduction, a particular scenario of interest involves \textit{local} doping/charged defect effects near surfaces. In such cases, CREST may be instrumental in compensating the excess charge introduced by such defects, allowing for a simultaneous treatment of both local and global doping effects. While in this work we have focused on electronic structures, we note that the derivation of the correct, full surface work function of arbitrarily charged surfaces can be obtained directly from the vacuum energy and the value of $E_F$ found by CREST. In addition, the correct calculation of doping-dependent total energies allows for the calculation of Hellman-Feynman forces and other derived quantities, and opens the door to the study of, e.g., surface phase diagrams and interface structures as a function of the bulk doping concentration.

\section*{Acknowledgements}
O. S. wishes to acknowledge helpful discussions with and input from Dr. Amir Natan (Tel Aviv University). This work was supported by the European Research Council, the Israel Science Foundation, the Helmsley Charitable Trust, the Lise Meitner Minerva Center for Quantum Chemistry, and the Deutsche Forschungsgemeinschaft (DFG) collaborative research project SFB 951 ``HIOS''. O. T. Hofmann acknowledges the FWF-project J 3285-N20. P. R. acknowledges the Academy of Finland through its Centres of Excellence Program (\# 251748).

\bibliographystyle{ws-procs975x65_Ofer}
\bibliography{crest-ms}

\end{document}